\documentclass[twocolappendix]{emulateapj}
\usepackage[breaklinks,colorlinks,allcolors=blue]{hyperref}
\usepackage{natbib}
\usepackage{graphicx}
\graphicspath{{figs/}}
\usepackage{CJK}

\usepackage{multirow, amsmath}
\begin{document}
\begin{CJK*}{UTF8}{gbsn}
\title{Blooming Trees: Substructures and Surrounding Groups of Galaxy Clusters}

\author{Heng Yu\altaffilmark{1,2,3}(余恒), Antonaldo Diaferio\altaffilmark{2,3}, Ana Laura Serra\altaffilmark{4}, 
Marco Baldi\altaffilmark{5,6,7}}

\altaffiltext{1}{Department of Astronomy, Beijing Normal University,
    Beijing, China 100875}
\altaffiltext{2}{Dipartimento di Fisica, Universit\`a di Torino, Via
  P. Giuria 1, I-10125 Torino, Italy}
\altaffiltext{3}{Istituto Nazionale di Fisica Nucleare (INFN), Sezione di Torino, Via
  P. Giuria 1, I-10125 Torino, Italy}
\altaffiltext{4}{On leave from the Dipartimento di Fisica, Universit\`a di Milano, Via Celoria 16, 20133 Milano, Italy}
\altaffiltext{5}{Dipartimento di Fisica e Astronomia, Alma Mater Studiorum Universit\`a di Bologna, 
                  via Gobetti 93/2, I-40129 Bologna, Italy;}
\altaffiltext{6}{INAF - Osservatorio Astronomico di Bologna, via Gobetti 93/3, I-40129 Bologna, Italy;}
\altaffiltext{7}{INFN - Sezione di Bologna, viale Berti Pichat 6/2, I-40127 Bologna, Italy;}


\begin{abstract}
We develop the Blooming Tree Algorithm, 
a new technique that uses spectroscopic redshift data alone to identify the substructures and the surrounding groups of galaxy clusters, along with their member galaxies. Based on the estimated binding energy of galaxy pairs, the algorithm
builds a binary tree that hierarchically arranges all the galaxies in the field of view.
The algorithm searches for buds, 
corresponding to gravitational potential minima on the binary tree branches; for each bud, the
algorithm combines
the number of galaxies, their velocity dispersion and their
average pairwise distance into a parameter that discriminates between the buds that do not
correspond to any substructure or group, and thus eventually die, and the buds 
that  correspond to substructures and groups, and thus bloom into the identified structures.
We test our new algorithm with 
a sample of 300 mock redshift surveys of clusters in different dynamical states; the clusters 
are extracted from a large cosmological $N$-body simulation of a $\Lambda$CDM model. 
We limit our analysis to substructures and surrounding groups identified in the simulation
with mass larger than $10^{13} h^{-1} M_{\odot}$.
With mock redshift surveys with 200 galaxies within 6 $h^{-1}$~Mpc from the cluster center, 
the technique recovers $ 80$\% of the real substructures and $ 60$\% of the surrounding groups; 
in $57$\% of the identified structures, at least 60\% of the member galaxies of the substructures and groups  
belong to the same real structure.
 These results improve by roughly a factor of two the performance of the best substructure identification algorithm
currently available, the $\sigma$ plateau algorithm, and  suggest that our Blooming Tree Algorithm can be an invaluable 
tool for detecting substructures of galaxy clusters and investigating their complex dynamics. 

\end{abstract}

\keywords{galaxies: clusters: general -- (cosmology:) large-scale
structure of universe -- methods: numerical, statistical}

\section{Introduction}
According to the standard cold dark matter paradigm, large cosmic structures form 
by merging 
of smaller structures \citep{1999Colberg,2005Colberg}. In this hierarchical universe, galaxy clusters 

form at later times, and, at the present time, some clusters are still accreting mass by 
merging. A clear signature of this process is the presence of substructures in the galaxy density distribution, in the X-ray and radio emission,
or in the dark matter distribution inferred from gravitational lensing \citep[e.g.,][and references therein]{2016Yu}.
Therefore investigating the properties of substructures can constrain the models of structure formation and evolution 
\citep{1982Geller,1996Mohr,2007Natarajan,2014Okabe,2016Mohammed, 2016Yu}, the connection between galaxy properties and environment 
\citep[e.g.,][]{2012Hwang,2013Pranger,2016Agulli,2016Utsumi,2017Oguri}, and even the nature of dark matter 
\citep[e.g.,][]{2015Harvey,2017Robertson,2017Kummer}.

Identifying dynamically distinct substructures in galaxy clusters is not a trivial task.
Most methods identify substructures in the galaxy density distribution based on spectroscopic data \citep[see][for a brief review]{2015Yu}. 
Among these methods, those relying on the hierarchical clustering
analysis appear to be particularly efficient. 

The hierarchical clustering analysis is a general statistical method.
It is designed to partition a system into optimally homogeneous subgroups 
on the basis of empirical measures of similarity \citep[see][for a detailed description]{everitt2011}.
\citet{1978Materne} first applied a hierarchical clustering analysis to astronomical data to identify groups of galaxies.
\citet{1996Serna} introduced the pairwise binding energy to link galaxies in the field of view of a cluster and
arrange them in a binary tree.
Building a binary tree
is a standard method to quantify the hierarchical structures of the entire system.
This approach does not rely on any morphological assumption or dynamical state, and it is thus suitable
for analyzing dynamically complex self-gravitating systems, like galaxy clusters.

In 1997, \citet{1997Diaferio} introduced the caustic method to estimate the mass profile of galaxy 
clusters in their outer regions, where the dynamical equilibrium assumption does not necessarily hold. 
In the detailed illustration of the algorithm of the caustic method, where galaxies are arranged
in a binary tree similarly to the procedure suggested by \citet{1996Serna}, \citet{1999Diaferio} first proposed 
the identification of a $\sigma$ plateau on the main branch of the tree 
to locate the cluster and return a list of cluster members and cluster substructures.
\citet{Serra2011} provide detailed and complete statistical tests of the caustic technique 
and propose a refined and more robust version of this $\sigma$ plateau algorithm.

The efficiency of the $\sigma$ plateau algorithm to identify the cluster substructures in $N$-body simulations 
is shown in \citet{2015Yu}, who 
emphasize a unique feature of this algorithm: unlike other methods 
for the identification of substructures with spectroscopic data, 
like the Dressler \& Shectman (DS) method \citep{1988DS,1999Solanes,2000Knebe,2010Aguerri,Dressler2013},
the KMM \citep{1994Bird,1996Colless,1998Barmby} or the DEDICA \citep{1996Pisani, 2007Ramella} algorithms, 
the $\sigma$ plateau algorithm gives an unambiguous association of galaxies 
to individual substructures; it thus enables the estimation of the substructure properties, 
like size, velocity dispersion and mass. 

With this feature available, we can apply the strictest criterion for comparing 
the substructures identified in three dimensions in an N-body simulation 
with the substructures identified in redshift space:  
in mock redshift surveys with 200 galaxies
within $3R_{200}$ from the cluster center, where $R_{200}$ is the usual radius of the sphere 
whose average density is 200 times the critical density of the Universe, the $\sigma$ plateau algorithm recovers $\sim 30-50$\% of the real substructures,
depending on the mass and the dynamical state of the cluster \citep{2015Yu}. 
This performance is unprecedented. The algorithm was successfully applied to the galaxy distribution in the field of
view of the cluster A85 \citep{2016Yu}: it provided a unique understanding of the complex dynamics of this cluster when combined
with the bulk motions of the intra-cluster medium in different regions, as inferred by the redshift measurements 
derived by X-ray spectroscopy.

Despite its very good performance, the $\sigma$ plateau algorithm actually overlooks the crucial fact that
cluster substructures can have widely different velocity dispersions.
Here, we present the Blooming Tree Algorithm, a new algorithm that takes this fact into account and thus represents a significant improvement over the $\sigma$ plateau algorithm. We show how this more sophisticated algorithm 
substantially doubles the substructure identification efficiency. In addition, the Blooming Tree Algorithm returns the list of the groups in the cluster outskirts along with their members. This feature provides a fundamental tool that enables a quantitative investigation of the merging and accretion history of galaxy clusters \citep{2001Rines,2013Lemze,2016DeBoni}.   

In Section \ref{sec:nbody}, we describe the
cosmological $N$-body simulation and the mock redshift surveys of the galaxy cluster fields
we use to test the method. We describe the Blooming Tree Algorithm and its results in Sections \ref{sec:tree} and \ref{sec:results}, 
respectively. 
In Section \ref{sec:caustic} we compare the performance of our new technique with the $\sigma$ plateau algorithm.
We conclude in Section \ref{sec:discussion}. 

\section{The Mock Redshift Catalogs of Simulated Clusters}
\label{sec:nbody}

We use the Coupled Dark Energy
Cosmological Simulations \citep[{\small CoDECS},][]{2012Baldi},  the largest set of
$N$-body simulations that model the interaction between a dark energy scalar field and the Cold
Dark Matter (CDM) fluid to date. Here, we only consider the simulation of the 
standard $\Lambda$CDM model with fiducial WMAP7 parameters \citep{2011WMAP}. The simulated volume is a
cube of 1 comoving $h^{-1}$~Gpc on a side
($ h = H_0 / 100$ km s$^{-1}$ Mpc$^{-1}$ is the dimensionless Hubble constant), containing $ 1024^3 $ CDM
particles with mass $5.84 \times 10^{10}h^{-1} M_{\odot} $
and the same number of baryonic particles with mass $1.17 \times 10^{10} h^{-1} M_{\odot} $.  
We only consider the dark matter particles: we assume 
that, in the real Universe, galaxies are unbiased tracers of the velocity
field of the dark matter particles. In fact, both $N$-body simulations 
\citep[e.g.][]{2001Diaferio,2004Gill,2004Diemand,2005Gill}
and observations \citep[e.g.][]{2008Rines,2016Rines} indicate that any velocity
bias between galaxies and dark matter is smaller than 10\%. 

Halos are identified with 
the Friends-of-Friends (FoF) algorithm \citep{1982FoF,1985Davis}, which links particles 
with distances less than the linking length $l_{\rm FoF}$ to form a group. 
We adopt the standard linking length $l_{\rm FoF} = 0.2 l_{\rm mean}$,
with $l_{\rm mean}$ the mean interparticle separation,
corresponding to the overdensity at virialization $\delta_v = \rho/\rho_{\rm b} = 185$ \citep{1998Audit},
with $\rho_{\rm b}$ the mean background density. In this procedure, the FoF halos 
are identified using the CDM particles as primary tracers and then linking baryonic 
particles to the group of their closest CDM neighbor. 

We also identify the three-dimensional (3D) substructures of the halos in the simulations
 with {\small SUBFIND} \citep{2001Springel}, whose algorithm is based on 
the overdensity and the gravitational binding energy of the particles
\citep[see][for further details]{2012Baldi}. 
With the mass of a 3D substructure we always indicate its total mass, namely the sum of the mass 
of the particles (both CDM and baryons) that are gravitationally bound to that substructure 
as identified by {\small SUBFIND}.

We consider a sample of 100 FoF halos 
at redshift $z=0$ within the mass range $10^{14}-10^{15} h^{-1} M_{\odot}$ 
with the aim of covering the variety of 
dynamical states; specifically, we consider 50 ``merging'' halos and 50 ``normal'' halos. 
We choose the 50 merging halos whose mass is closest to $5\times 10^{14} h^{-1} M_{\odot}$ and that 
contain a substructure whose mass is at least half the mass
of the halo core, where the core is the SUBFIND substructure whose center coincides with
the halo center.  Among the remaining halos between $10^{14}-10^{15} h^{-1} M_{\odot}$, we
select the 50 normal halos whose mass is closest to $5\times 10^{14} h^{-1} M_{\odot}$ and
that are not merging halos.  The masses of our 100 halos
are in the range $4.17-6.39 \times 10^{14} h^{-1} M_{\odot}$, with a median mass 
$4.93 \times 10^{14} h^{-1} M_{\odot}$. 

We locate each halo at the center of the simulation box 
exploiting the periodic boundary conditions.
To mimic the observation of real clusters, we assign the celestial coordinates   
$(\alpha,\delta)=(6^h,0^\circ)$ and a redshift distance $cz=36000$~km~s$^{-1}$
to the halo center.
Around the halo, we consider a rectangular prism enclosing the volume 
corresponding to a solid angle that at the halo
distance ensures to cover a square area $12 h^{-1}$~Mpc wide. The volume
is centered on the halo and it is $140 h^{-1}$~Mpc deep.
The resulting field of view (FoV) is $1.6^{\circ} \times 1.6^{\circ}$.
For each halo, we apply this procedure to three orthogonal directions.
Since the halos are generally not spherically symmetric, for our statistical purposes 
we can consider these three catalogs as independent mock clusters.
So we obtain 300 mock redshift catalogs.

The observational volumes we extract from the simulation 
typically contain $\sim 6 \times 10^4$ particles.
To get a number of particles close to a realistic number of observable galaxies,
we randomly sample the dark matter particles until we obtain a given number of particles $N_c$ 
within a sphere of radius 6 $h^{-1}$~Mpc from the halo center, corresponding to $\sim 5 R_{200}$ for the halos of 
our sample. This procedure yields mock catalogs with different
numbers of total particles $N$. 
To explore the effect of particle sampling, we build catalogs 
with $N_c=$~(50, 100, 150, 200, 250, 300). Additionally, we only retain particles in the
mock catalogs whose redshift is within 
$\pm 4000$~km~s$^{-1}$ from the halo redshift.

\begin{figure}
\includegraphics[width=.45\textwidth]{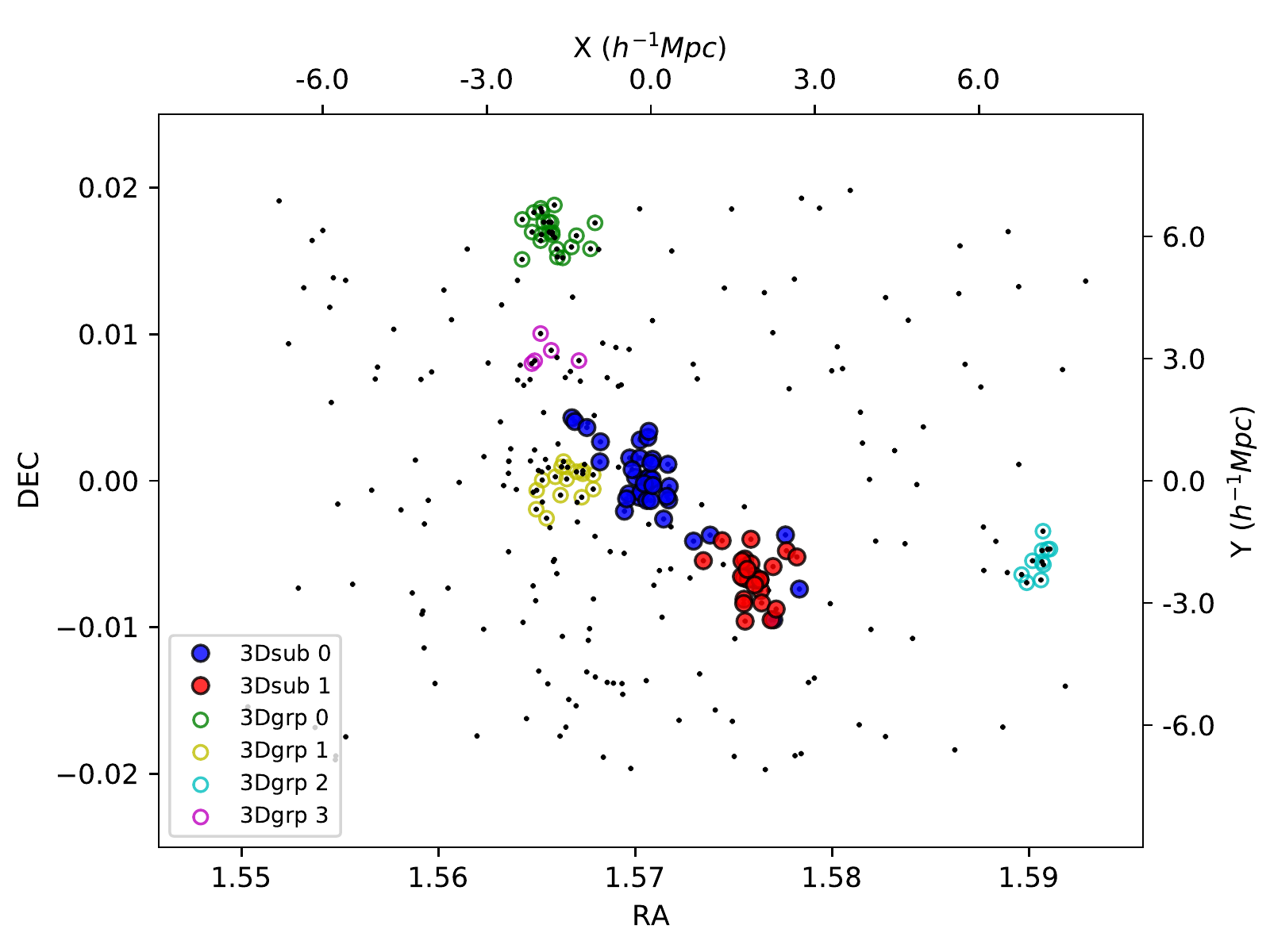}\\
\includegraphics[width=.45\textwidth]{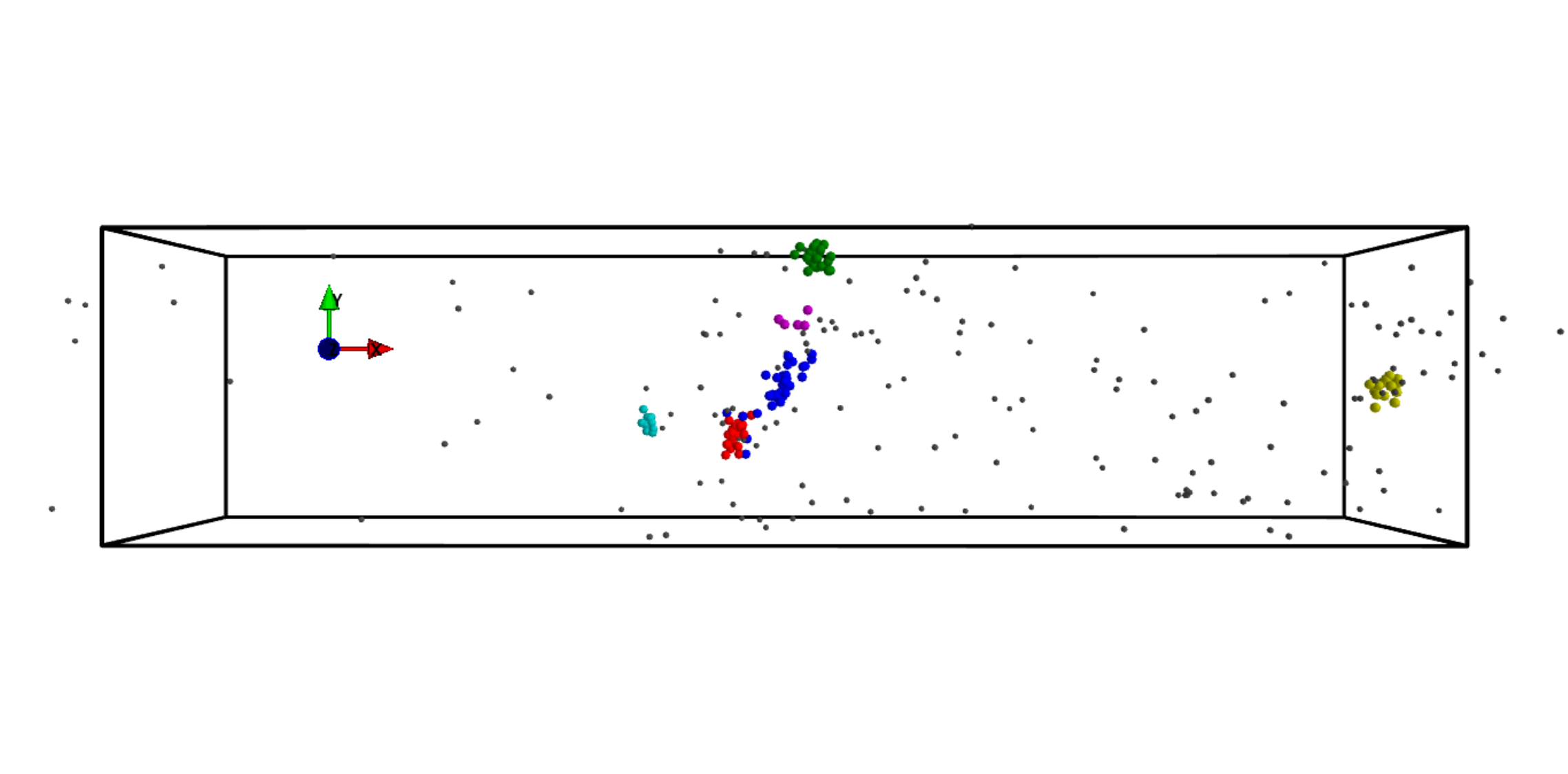}
\caption{The upper panel shows the distribution of the 3D substructures and the surrounding groups of a  merging 
halo on the plane of the sky. 
The axis labels show both the projected celestial coordinates RA and DEC, in radiants, and the 
comoving coordinates in the $N$-body simulation. 
The blue solid dots show the particles in the halo core at the center of the FoV (sub 0 in the inserted legend).
The red solid dots show the members of the largest 3D substructure that identifies this halo
as a merging halo (sub 1 in the legend). 
The colored open circles show the members  of the surrounding groups,
as listed in the inserted legend. The remaining black points show the remaining particles in the FoV.
The lower panel provides a three-dimensional perspective of the system. We overplot a box with
dimensions $14 \times 14 \times 60 h^{-3}$~Mpc$^3$.} 
\label{fig:realsky}
\end{figure}

By randomly sampling the dark matter particles, 
the number of members of a 3D substructure in the mock catalog 
can be substantially reduced or even vanish.  
We only consider 3D substructures that have at least 10 particles appearing in the
FoV.  An example is shown in Figure \ref{fig:realsky}.
It is a mock catalog with $N_c=100$ and $N=304$.

The properties of our mock catalogs are listed in Table \ref{table:fields}.
The third to fifth columns list the medians and percentile ranges of the number of 
particles $N$ in the FoV as a function of $N_c$.
These mock redshift surveys are comparable 
to recent large galaxy surveys of clusters and their surroundings, 
such as CIRS \citep{Rines2006} and HeCS \citep{Rines2013}.
The seventh column is the total number of 3D substructures $n_{\rm sub}$,
with at least 10 member particles in the FoV and with mass larger than $ 10^{13} h^{-1} M_\odot$. 
This threshold is the minimum 3D substructure mass set by the number of luminous galaxies 
that can be detected in current typical surveys: a $10^{13} h^{-1} M_\odot$ substructure is 
expected to contain a handful of galaxies brighter than $10^{10} h^{-1} L_\odot$ at most. 
Table \ref{table:fields} also lists the total number of halos 
$n_{\rm cl}$ and the total number of groups $n_{\rm grp}$,  namely the 3D halos found by the FoF algorithm, that surround each individual central halos.
As expected, the number of the 3D surrounding groups appearing in the FoV increases with increasing $N_c$,
whereas the number of 3D substructures appears to be less sensitive to $N_c$.
Hereafter we refer to the halo at the center of the FoV as the {\it cluster} and to the particles in the FoV as the 
{\it galaxies}. 

The Blooming Tree Algorithm identifies three different kinds of structures: 
(1) substructures; (2) cores; and (3) surrounding groups.
The substructures only contain
main cluster members, namely the galaxies linked by the FoF algorithm;
the core is the substructure that contains the cluster center; 
the surrounding groups are structures containing galaxies that are not members of the
main cluster. 
Below we generically indicate any of these three kinds of structures as {\it structures}, unless specified otherwise.

\begin{table}
\caption{The number of particles $N$ and structures in the FoV.}
\centering
\begin{tabular}{|c|c|lll|lll|}
\hline
\multirow{2}{*}{$N_{c}$} &\multirow{2}{*}{type} &\multicolumn{3}{|c|}{$N$} & \multirow{2}{*}{$n_{\rm cl}$} & \multirow{2}{*}{$n_{\rm sub}$} & \multirow{2}{*}{$n_{\rm grp}$} \\
\cline{3-5}
    &  & 10\% & 50\% & 90\% &  &  &  \\
\hline
\multirow{3}{*}{50}  & normal & 127 & 176 & 257 & 150 & 21 & 233 \\
 & merging & 128 & 173 & 268 & 150 & 133 & 274 \\
 \cline{2-8}
 & total & 128 & 176 & 264 & 300 & 154 & 507 \\
 \hline
\multirow{3}{*}{100} & normal & 250 & 362 & 502 & 150 & 48 & 532 \\
 & merging & 258 & 364 & 523 & 150 & 185 & 550 \\
 \cline{2-8}
 & total & 256 & 364 & 515 & 300 & 233 & 1082 \\
 \hline
\multirow{3}{*}{150} & normal & 400 & 527 & 747 & 150 & 73 & 821 \\
 & merging & 397 & 526 & 800 & 150 & 217 & 855 \\
 \cline{2-8}
  & total & 399 & 527 & 776 & 300 & 290 & 1676 \\
  \hline
\multirow{3}{*}{200}  & normal & 522 & 709 & 1004 & 150 & 83 & 1090 \\
 & merging & 506 & 717 & 1048 & 150 & 244 & 1144 \\
 \cline{2-8}
  & total & 515 & 716 & 1010 & 300 & 327 & 2234 \\
  \hline
\multirow{3}{*}{250}  & normal & 653 & 882 & 1279 & 150 & 82 & 1357 \\
 & merging & 649 & 889 & 1268 & 150 & 255 & 1391 \\
 \cline{2-8}
  & total & 649 & 886 & 1279 & 300 & 337 & 2748 \\
  \hline
\multirow{3}{*}{300}  & normal & 795 & 1044 & 1571 & 150 & 102 & 1592 \\
 & merging & 783 & 1066 & 1528 & 150 & 276 & 1595 \\
 \cline{2-8}
  & total & 795 & 1063 & 1562 & 300 & 378 & 3187 \\
  \hline
\end{tabular}
\label{table:fields}
\end{table}

\section{The Blooming Tree Algorithm}
\label{sec:tree}

\subsection{Tree Construction}
According to hierarchical clustering models, clusters of galaxies form by the
aggregation of smaller systems accreting from the surrounding region. 
To good approximation, galaxies are collisionless objects during cluster merging, 
and the transfer of mechanical energy to galaxy internal degrees of freedom is negligible.
If the 3D binding energy can be fairly represented by their 
projected values, 
we can infer the internal structures of a cluster 
based on a hierarchical clustering analysis where we adopt the galaxy projected binding energy 
as the similarity. 
Although the 3D and projected binding energies of an individual pair might be largely discrepant from each other, these two quantities 
in a sample of pairs are strongly correlated, supporting our adoption of the projected binding energy in the hierarchical clustering analysis. \footnote{Non-parametric statistical tests between the 3D binding energy of individual galaxy pairs in our simulation and their two-dimensional (2D) binding energy estimated with equation (\ref{eq:pairwise-energy}) demonstrate that the correlation between the two quantities is strong: for our sample of $\sim 1.4\times 10^7$ pairs, 
we find the Spearman and Kendall 
rank correlation coefficients $r=0.50$ and $\tau=0.46$, respectively, which have a probability $P<10^{-30}$ to appear
for an uncorrelated sample. These coefficients increase to $r=0.79$ and $\tau=0.62$, again with probability $P<10^{-30}$, 
if we limit the sample to bound pairs, namely pairs with negative 3D binding energy. 
To find a probability $P>10^{-30}$, we need to take a random subsample of less than 100 pairs: for a subsample of 36 pairs, we find
$P=4\times 10^{-8}$ and $P=4\times 10^{-9}$, for the  Spearman and Kendall
correlation coefficients, respectively, showing that the correlation remains robust even for relatively small pair samples.}

We perform the hierarchical clustering analysis by building a binary tree as follows 
(see \citealt{1999Diaferio} and \citealt{Serra2011} for further details): 

\begin{itemize}
 \item[i.] initially each galaxy is a group $g_\alpha$;
 \item[ii.] the binding energy $E_{\alpha\beta}={\rm min}\{E_{ij}\}$, where $E_{ij}$ 
is a projected binding energy between the galaxy $i\in g_\alpha$ and the
galaxy $j\in g_\beta$, is the similarity associated to each group pair $g_\alpha, g_\beta$. 
The projected binding energy is estimated with the equation 
\begin{equation}
E_{ij}=-G{m_i m_j\over R_p}+{1\over 2}{m_i m_j\over m_i+m_j}\Pi^2 \;,
\label{eq:pairwise-energy}
\end{equation}
where $R_p$ is the pair projected separation, $\Pi$ is the
line-of-sight velocity difference and $m_i=m_j=10^{12}h^{-1}$ 
M$_\odot$ is the typical total mass of a luminous galaxy;\footnote{ At this stage, we refrain from including 
different masses for galaxies of different luminosity to avoid the introduction of additional degrees of freedom: the mass-to-light ratio depends on the galaxy morphological type and luminosity, which, on turn, depends on the fixed angular aperture generally used for the photometric measurement; therefore, including the connection between the measured luminosity and the galaxy mass requires a not negligible number of parameters. } 

\item[iii.] the two groups with the smallest binding energy $E_{\alpha\beta}$ are
replaced with a single group $g_\gamma$ and the total number of groups is decreased by one;
\item[iv.] the procedure is repeated from step (ii) until only one group is left.

\end{itemize}

\begin{figure*}
\centering
\includegraphics[width=.80\textwidth]{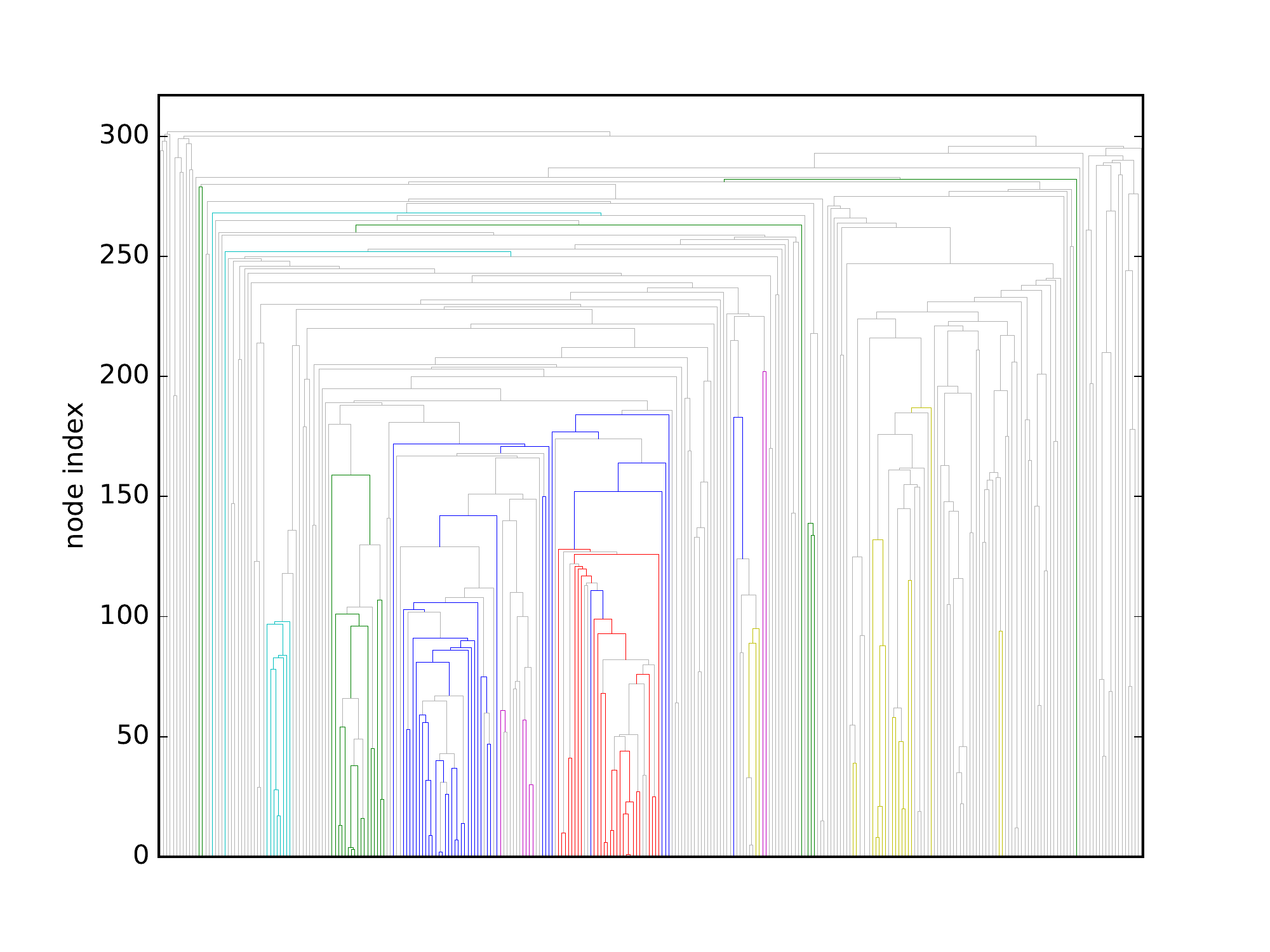}\\
\includegraphics[width=.80\textwidth]{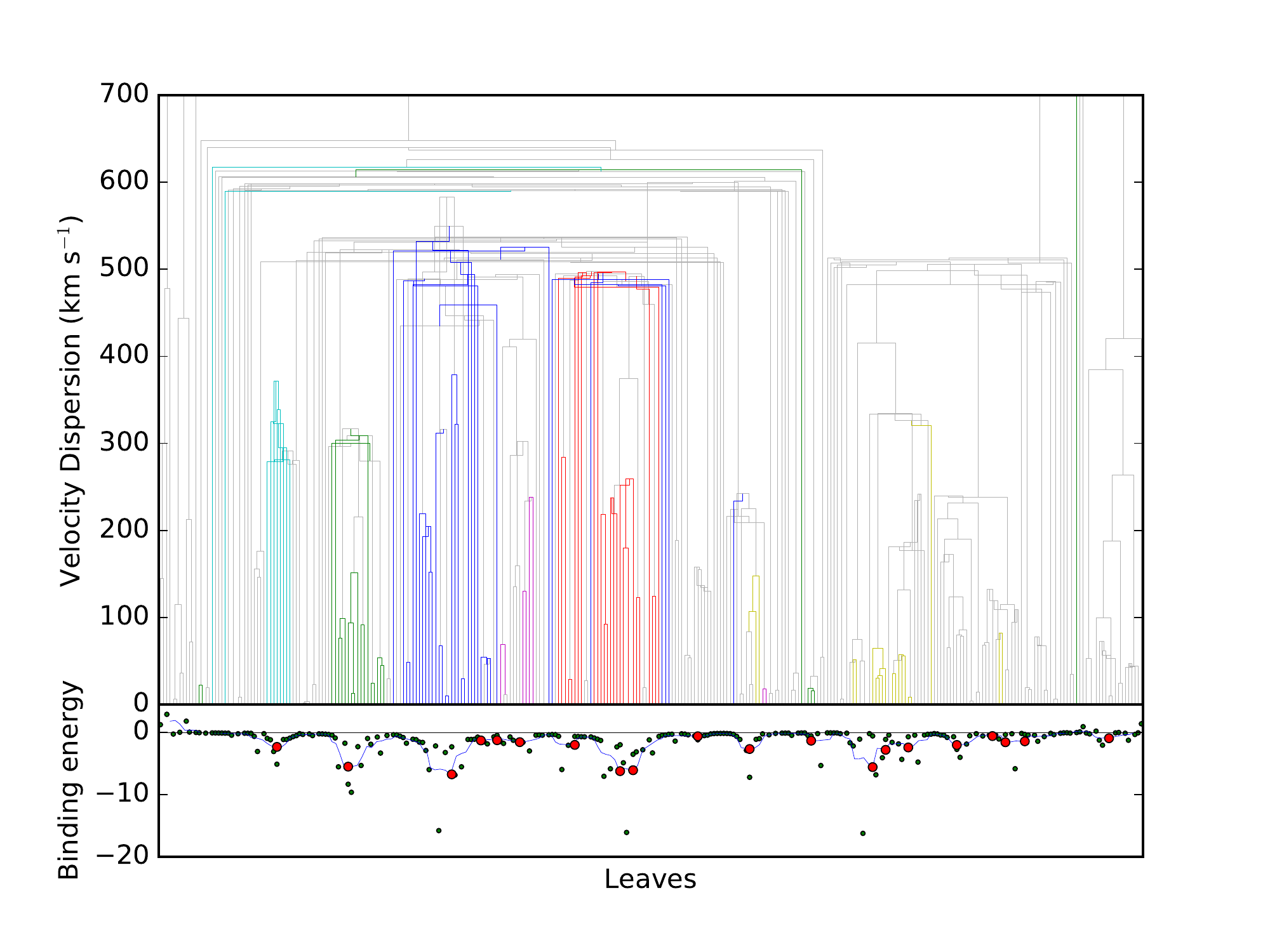}
\caption{Two dendrograms illustrating the binary tree of the merging cluster shown in Figure
\ref{fig:realsky} with 304 galaxies in the FoV. 
The galaxies are the leaves of the tree at the bottom of each dendrogram. 
All galaxies belonging to different structures are shown
with different colors, with the same color code of Figure \ref{fig:realsky}.
In the top panel, the $y$-axis of the dendrogram displays the node index: larger indexes indicate less bound branches.
In the bottom panel, the $y$-axis of the dendrogram displays the node velocity dispersion.
The black dots, one for each leave, in the lower inset in the bottom panel show the binding energy profile of the binary tree: 
the blue solid curve is obtained by smoothing the profile with a 5-leaf wide box filter; 
the red dots show the local minima, or buds, whereas the horizontal solid black line indicates binding energy equal to zero.
} 
\label{fig:dendrogram}
\end{figure*}

At this stage all the galaxies, now the leaves, are arranged in a binary tree,
that quantifies their hierarchical relationship.
When plotting the dendrogram representing the binary tree (Figure \ref{fig:dendrogram}, top panel), the node index, namely the similarity, 
or the binding energy in our case, is adopted as the quantity shown on the vertical axis \citep{1996Serna}.
However, in our case, the vertical axis can show more information when a different quantity is displayed.

Here we choose to show the velocity dispersion of each node,  
estimated with all the leaves hanging from that node, because it
is a physical property directly related to the depth
of the gravitational potential well of a bound structure.
The velocity dispersion of the nodes 
is not always a monotonic function when walking from the root to the leaves;
therefore, unlike the projected binding energy,  
displaying the velocity dispersion on the $y$-axis of the dendrogram might produce
branches that intersect each other, as shown in the bottom panel of Figure \ref{fig:dendrogram}.
However, this choice is more advantageous than displaying the similarity, 
because it generally separates different bound structures 
more  clearly. 

The dendrograms shown in Figure \ref{fig:dendrogram} correspond to 
the merging cluster in the FoV shown in 
Figure \ref{fig:realsky}:
the member galaxies of the core (in blue) and of the main substructure (in red)
are mainly separated into two branches. However, 
due to projection effects, the members of the same structure 
are not always close to each other on the dendrogram.
Surrounding groups (with green, yellow, cyan and magenta circles, in Figure \ref{fig:realsky}) 
also appear as distinct branches of the binary tree.
The goal of the next step of our Blooming Tree Algorithm is 
to identify the branches corresponding to these 
structures.

Incidentally, we remark here that in the language of the standard cluster analysis \citep[e.g.][]{everitt2011,hennig2015}, our binary tree construction 
is based on a single-linkage hierarchical algorithm. Being based on an estimate of the pairwise gravitational binding energy of galaxies, which is the fundamental physical quantity for identifying gravitational structures, our approach is the best physically motivated method, and
compensates known shortcomings of the single-linkage method, like the tendency of connecting independent structures, similarly to 
the standard friends-of-friends algorithm \citep{1982FoF}. The physical interpretation of the linkage in other standard 
agglomerative methods, less prone to this shortcoming, would necessarily be, in this context, more vague and questionable.

\subsection{Buds: Binding Energy Minima}

Substructures hang from different nodes of separated branches.
They may have different velocity dispersions and different binding energies
because of different masses. The identification of
these structures requires the identification of the
proper branches of the binary tree.

To locate the minima of the gravitational potential wells, we consider the 
binding energy of all the leaves. An example is shown in the lower inset in the bottom panel of
Figure \ref{fig:dendrogram}.
This binding energy shows deep minima corresponding to real structures, 
and fluctuations mostly caused by chance alignments. To suppress this noise,
we smooth the profile with a box filter that is 5-leaf wide.
The  blue solid curve in the  inset shows the smoothed profile.

We call buds the local minima of this smoothed binding energy profile. 
Smoothing algorithms more sophisticated than the box filter are clearly possible, but they are unnecessary for our only purpose here of finding the buds from the local minima of the binding energy curve. 
The buds identify the branches that will be
searched for the identification of the real structures, as illustrated 
in the next step.

\subsection{Branch Searching and Blooming Buds }
\label{sec:branch}

Let us consider the dendrogram in the bottom panel of Figure \ref{fig:dendrogram}, where the vertical axis shows the velocity dispersion
of each tree node, and let us walk from the leaves to the root  on a given branch by moving from one node to its parent node. Being 
a binary tree, this path implies adding a leaf at each step. We see that the velocity dispersions of the nodes on the same branch either
are basically unaffected by the step towards the root or they vary subtantially. In other words, when moving from the leaves to the root, 
the velocity dispersion on a given
branch does not generally grow smoothly, but it shows either sudden jumps or plateaus.

These plateaus indicate the presence of structures.
The original version of the $\sigma$ plateau algorithm identifies one single plateau 
on the main branch of the tree and isolates the cluster and its substructures \citep{1999Diaferio,Serra2011,2015Yu}.
In systems with complex dynamics, the single plateau might not actually be flat or there might 
be several plateaus; identifying a single plateau is thus not obvious: 
a too large threshold may erroneously associate distinct substructures into a single substructure, 
whereas a too small threshold may separate an individual substructure into smaller clumps.

Here, we implement a new algorithm that combines
three pieces of information which are missing from the original version of the  $\sigma$ plateau algorithm:
the line of sight velocity dispersion $\sigma_v$ of the leaves hanging from a node, the number
$n$ of these leaves, and the distribution of the leaves on the sky;
$\sigma_v$ and $n$ are combined in the average velocity dispersion $\sigma_v/n$, because, 
when two branches corresponding to two distinct structures, 
with similar $\sigma_v$ and $n$, merge,  
the branch of the combined structure has $\sigma_v$ comparable to the
original structures, but $\sigma_v/n$ is reduced by a factor $\sim 2$.
Therefore, the relation $\sigma_v$ vs. node index is roughly flat, 
whereas the relation $\sigma_v/n$ vs. node index shows a rapid decrease.
We also take the distribution of the galaxies on the sky into account, because this piece of information 
is crucial to identify a real 3D 
structure, as we show below.

To implement these three diagnostics, we consider the mass of a system estimated 
with the virial theorem
\begin{equation}
M_v = {4 \over 3} \pi R_v^3 \delta_v \rho_c = \alpha {3\sigma_v^2 R_v \over G}
\label{eq:mtot}
\end{equation}
where $R_v$ is the virial radius,  
the average density of the system is $\delta_v$ times 
the critical density $\rho_c = 3H^2/8 \pi G$, with 
$H$ the Hubble parameter, and $\alpha$ is a numerical factor of order unity that depends
on the mass distribution within the system; hereafter, we neglect this constant, 
because it is irrelevant for our purpose.
All self-gravitating systems in dynamical equilibrium thus satisfy the relation 
\begin{equation}
{\sigma_v \over R_v} = \sqrt{{4 \over 9} \pi G \delta_v \rho_c} = \sqrt{\delta_v \over 6} H \;,
\label{eq:vRv}
\end{equation}
unlike random associations of unrelated galaxies. 
The trend of $\sigma_v/R_v$ against the node number will thus show a discontinuity 
when walking on the binary tree from a bound structure to a region containing unrelated galaxies. 
As a robust estimate of $R_v$, we use the two-dimensional average distance $d_{avg} = 2 R_v / n^{1/2} $, with
\begin{equation}
d_{avg} = {{\Sigma_{i \neq j} \sqrt{(x_i - x_j)^2 + (y_i - y_j)^2}} \over n(n-1)} \; ,
\label{eq:davg}
\end{equation}
where $x_{i,j}$ and $y_{i,j}$ are the Cartesian coordinates of the galaxies $i$ and $j$ on the
plane of the sky.   

\begin{figure}[htbp]
\includegraphics[width=0.45\textwidth]{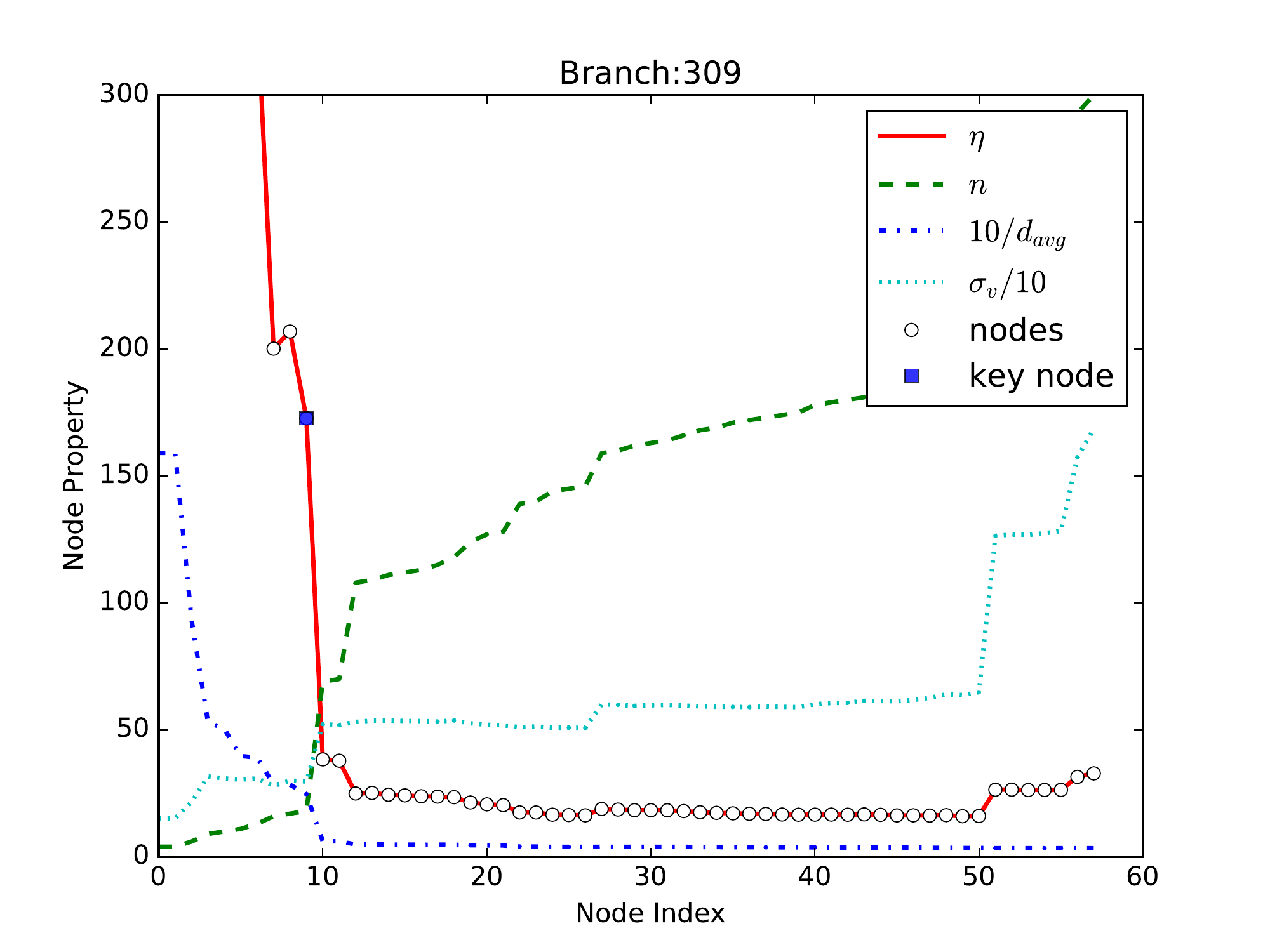}
\caption{The node properties of one branch (grp 0, green) of the binary tree 
of Figure \ref{fig:dendrogram} from the leaves to the root; 
$d_{avg}$ and $\sigma_v$ are rescaled to fit into the plot.
The blue square indicates the key node where we cut the branch.}
\label{fig:avel}
\end{figure}

By replacing $R_v$ in equation (\ref{eq:vRv}) with $d_{avg}$ and dropping the
factor 2, which is irrelevant for our purpose, we finally adopt the following quantity to identify the structures
\begin{equation}
\eta = {\sigma_v \over n^{1/2} \, d_{avg}}  \; {\rm km\, s^{-1}\, Mpc^{-1}} \; .
\label{eq:equivalent}
\end{equation}
Figure \ref{fig:avel} shows the typical trends of some of the quantities defined
above with the node index; $\sigma_v$ and $d_{avg}$ 
have been  arbitrarily rescaled to fit into the plot.
The figure shows that all the three quantities, the velocity dispersion $\sigma_v$, 
the galaxy number $n$, 
and the average distance $d_{avg}$, increase when the branch includes more galaxies.
In passing, we note that $\sigma_v$ has an upper limit, outside the
range shown in Figure \ref{fig:avel}, because we limit the redshift 
range of the sample to $\pm 4000$~km~s$^{-1}$. 
Also, a small $d_{avg}$ makes $\eta$ very large for compact systems. 

When, by walking along a branch corresponding to a real structure, we start including interloping galaxies, 
the increased 
number of galaxies and the increased average distance are not
compensated by a proportional increase of $\sigma_v$: $\eta$ will thus 
abruptly decrease with the node index. 
This downward jump of $\eta$ can be used as a diagnostic for the 
identification of the structures. However, the amount of the decrease can 
substantially vary from case to case, depending on physical conditions, like local density, richness and mass of 
the structure, and observational constraints, like completeness and survey density. A more suitable diagnostic is 
$\Delta \eta $, the difference between the value of $\eta_{sub}$ associated
to the structure and the value of $\eta_{bck}$ associated to the background region
surrounding the structure.

We define $\eta_{sub}$ as the value of $\eta$ 
for each 3D structure with at least ten galaxies 
in the FoV, and $\eta_{bck}$ as the value of $\eta$ of the region centered on each 
3D structure and extending to a projected radius $R_5$, which is five times larger than the radius of the structure,
estimated by the radius of the
smallest circle enclosing the structure on the plane of the sky. To estimate $\eta_{bck}$, we 
use equation (\ref{eq:equivalent}), where now $n$ is the number of galaxies within the circle of radius $R_5$ and 
$\sigma_v$ is their velocity dispersion. This definition of $\eta_{bck}$ quantifies how the value of $\eta$
of a 3D structure is affected by the inclusion of galaxies appearing in the surrounding area projected on
the sky, which are likely to be unbound to the structure. This definition is more appropriate than, for example, choosing 
a random area in the field of view in the neighborhood of the structure, because the galaxy distribution is 
highly inhomogeneous on these cosmic scales, and the probability of selecting another structure in the random field is not negligible; 
the value of  $\eta_{bck}$ would thus be not representative of a random sample of unrelated galaxies.

\begin{figure}
\includegraphics[width=0.45\textwidth]{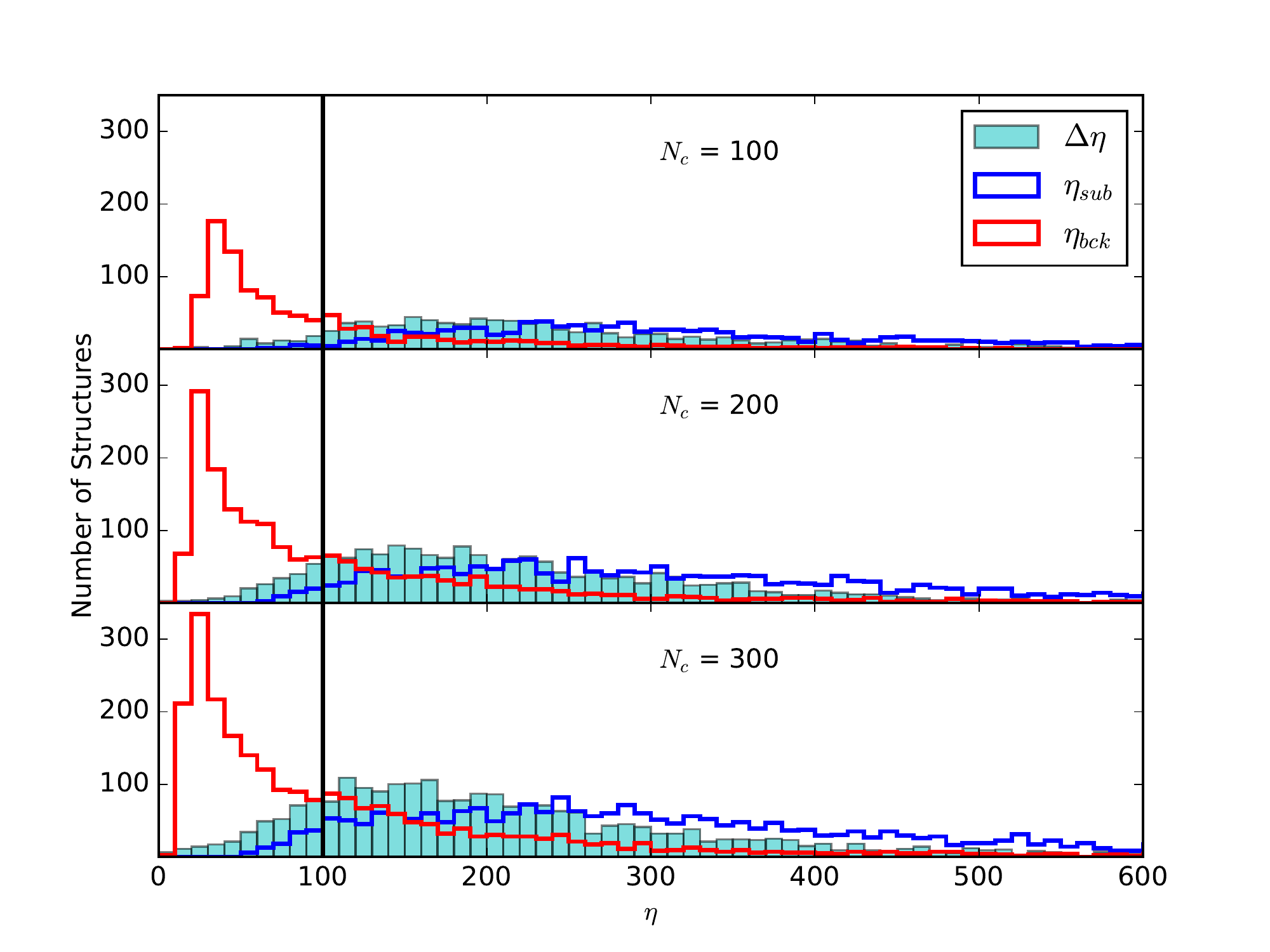}
\caption{The distribution of $\eta_{sub}$ from all the 3D structures with 
at least ten particles (open blue histogram), 
the distribution of $\eta_{bck}$ from the region centered on each structure and extending to 
a radius five times larger than the radius of the 3D structure (open red histogram), and 
the distribution of $\Delta \eta = \eta_{sub}-\eta_{bck}$ (solid cyan histogram). 
The vertical line indicates our threshold $\Delta \eta = 100$. Each panel is for a
different number $N_c$ of member galaxies in the cluster, as indicated in each panel.  }
\label{fig:hst_dh}
\end{figure}

Figure \ref{fig:hst_dh} shows that the distribution of $\eta_{sub}$, the blue open histogram, is rather flat and mostly extends beyond $\eta_{sub}= 100$;
on the contrary, most $\eta_{bck}$, the open red histogram, tend to be closer to zero,  because $d_{avg}$ increases
more rapidly than $n$ and $\sigma_v$. The distribution of $\Delta \eta=\eta_{sub} -\eta_{bck}$, the solid histogram, qualitatively 
resembles the distribution of $\eta_{sub}$.

The three panels in Figure \ref{fig:hst_dh} show the distributions of $\eta_{sub}$, $\eta_{bck}$, and $\Delta \eta$, for
three values of the $N_c$ member galaxies in the cluster, to mimic different densities of the redshift survey. 
 The value of $N_c$  slightly affects these distributions: 
  the 10th, 50th, and 90th percentiles of the  $\eta_{sub}$ distributions 
 are $(160, 298, 574)$, $(139, 297, 600)$, and $(127, 288, 608)$  for $N_c=100$, $200$, and $300$, respectively; 
similarly, the percentiles for the $\eta_{bck}$ distributions
are $(32, 66, 227)$, $(24, 70, 260)$, and $(21, 73, 286)$, and for $\Delta \eta$ are   
$(103, 213, 415)$, $(89, 195, 404)$, and $(69, 183, 400)$. To filter out random associations of unrelated galaxies and
identify the list of structure candidates, 
we adopt a threshold for  $\Delta \eta$. In Sect. \ref{sec:SRandC} below,
we define the success rate and completeness of the sample of two-dimensional (2D) structures identified
with the Blooming Tree Algorithm; we show how these two quantities vary with 
the threshold $\Delta \eta$. For $\Delta \eta=100$, which is close to the 10th percentile
of the $\Delta \eta$ distribution, the completeness is maximized. We thus choose this threshold
for our following analysis. We have also tested that other values of this threshold, 
in the range $\Delta \eta=(50,120)$, leave our results substantially unaffected.

Given the similarity of the trends of $1/d_{avg}$ and $\eta$ shown in Figure \ref{fig:avel}, 
we might believe that the same argument used for $\eta$ could be 
applied to $1/d_{avg}$, and that $d_{avg}$, rather than $\eta$, could be used as a diagnostic. This
is not the case, however, because the two parameters $\sigma_v$ and $n$, contained in $\eta$, partly balance the variation of $d_{avg}$
with different densities of the redshift survey and different densities of the system itself: 
$d_{avg}$ varies from compact to loose groups, that  are both
real systems, and it is different in the center and in the outer regions of clusters. Therefore, a single threshold on $d_{avg}$, or $\Delta d_{avg}$, is unable, unlike
 $\Delta \eta$, to identify 
structures in different environments and with different observational constraints.

To identify the 2D structures from the redshift data alone, the algorithm proceeds as follows: 
we explore all the branches  showing a bud, or a local 
minimum of the binding energy profile. 
We compute $\Delta \eta $ along each branch from the leaves to the root, and we
label as a  key node the node before the last downward variation larger
than the threshold $\Delta \eta = 100$. 
 The key node from which at least  6 galaxies hang and no other
key node hangs identifies a 2D structure: in other
words, the bud associated to the branch containing this key node becomes a flower and our tree blooms. 

\begin{figure*}[htbp]
\includegraphics[width=0.95\textwidth]{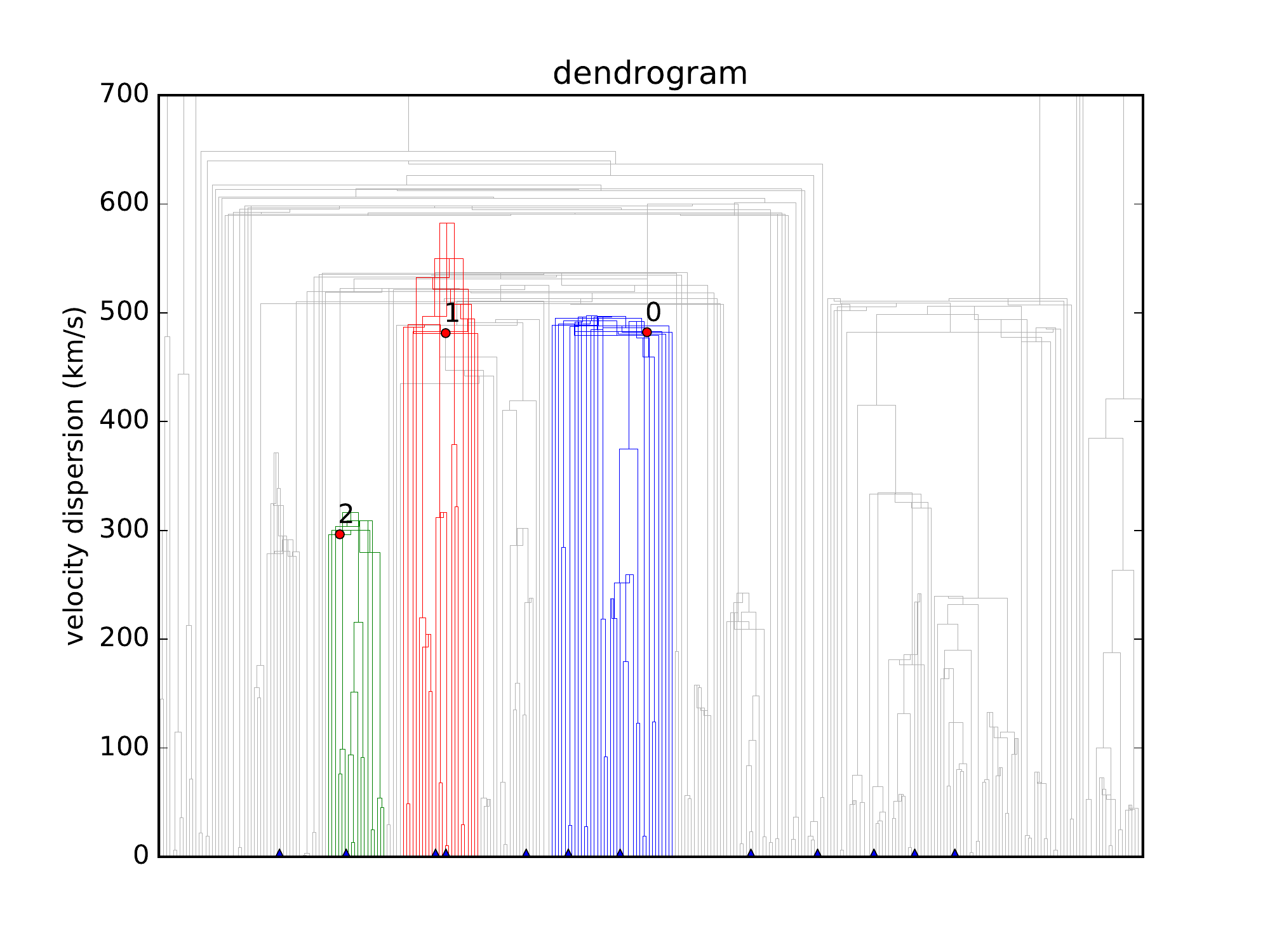}
\caption{The dendrogram displayed with the velocity dispersion of every nodes on the $y$-axis. 
All the branches having a bud, or a local minimum of the binding energy, are labeled with blue triangles at the bottom of the panel. 
All the key nodes found with the $\Delta \eta$ threshold, that have  bloomed, are highlighted with red dots.
The stems of their member leaves are indicated with different colors.}
\label{fig:vtree}
\end{figure*}

\begin{figure}[ht]
\includegraphics[width=0.45\textwidth]{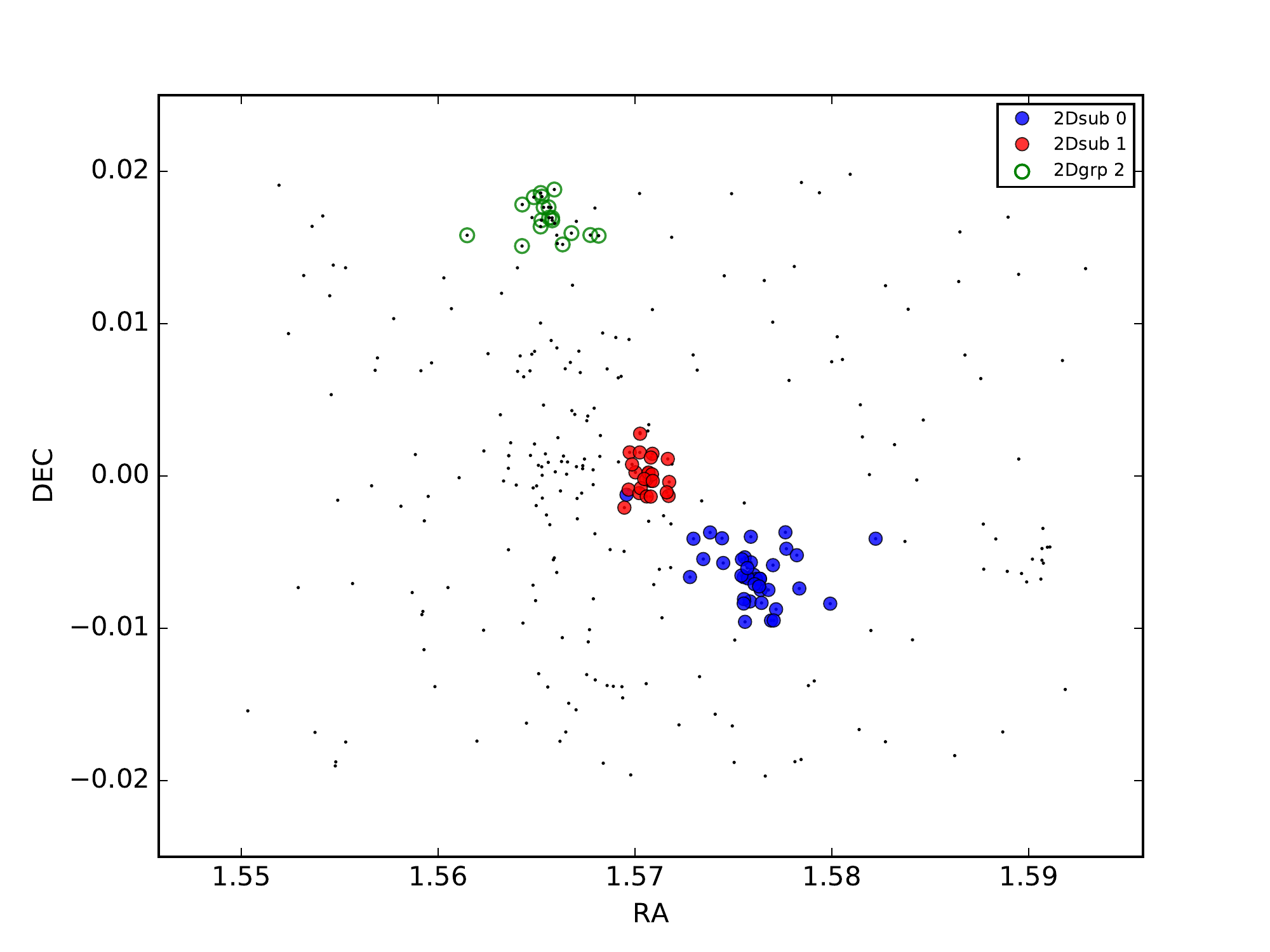}
\caption{The distribution on the sky of the 2D structures identified with the Blooming Tree Algorithm. 
The color code is the same as Figure \ref{fig:vtree}.}
\label{fig:subs}
\end{figure}

Figure \ref{fig:vtree} shows an example of this binary tree analysis for the same system
shown in Figure \ref{fig:realsky}. 
The corresponding distribution on the sky of the identified structures is 
shown in Figure \ref{fig:subs}:
visually, the rich substructures and groups appear to be recovered 
at the proper position and with 
their proper size. In the next section, we compare the properties of the
2D structures with the 3D structures and provide a statistical 
analysis of the performance of our structure identification method.

\section{Performance of the blooming tree algorithm}
\label{sec:results}

\subsection{Success Rate and Completeness}
\label{sec:SRandC}

The results shown in this subsection are for the
full sample of 50 merging clusters and 50 normal clusters projected along three orthogonal lines
of sight and sampled with six different values of $N_c$ for a total number of $(50+50)\times 3\times 6=1800$ mock
catalogues. We show the dependence of these results on the cluster dynamical state and on the FoV sampling in Sect. \ref{sec:statcomp}.
  
To quantify whether a 2D structure corresponds to a 3D structure, 
we make a one-to-one comparison 
between the members of the 2D structures identified with the Blooming Tree Algorithm and the members of
the 3D structures identified with {\small SUBFIND}. The possibility of this
one-to-one comparison is unique to the $\sigma$ plateau algorithm and the Blooming Tree Algorithm. 

A single 2D structure may contain members belonging to different 3D structures or none. 

For each 2D structure, we compute 
\begin{equation}
 f_{3D} = \frac{n (mem_{3D} \in mem_{2D}) }{n_{2D}}
\end{equation}
where $n_{2D}$ is the total number of members of the 2D structure and $n (mem_{3D} \in mem_{2D})$ is the largest number of particles, among the $n_{2D}$ members of the 2D structure, that
are also members of a single 3D structure with $n_{3D}$ members.

\begin{figure}[htbp]
\includegraphics[width=0.45\textwidth]{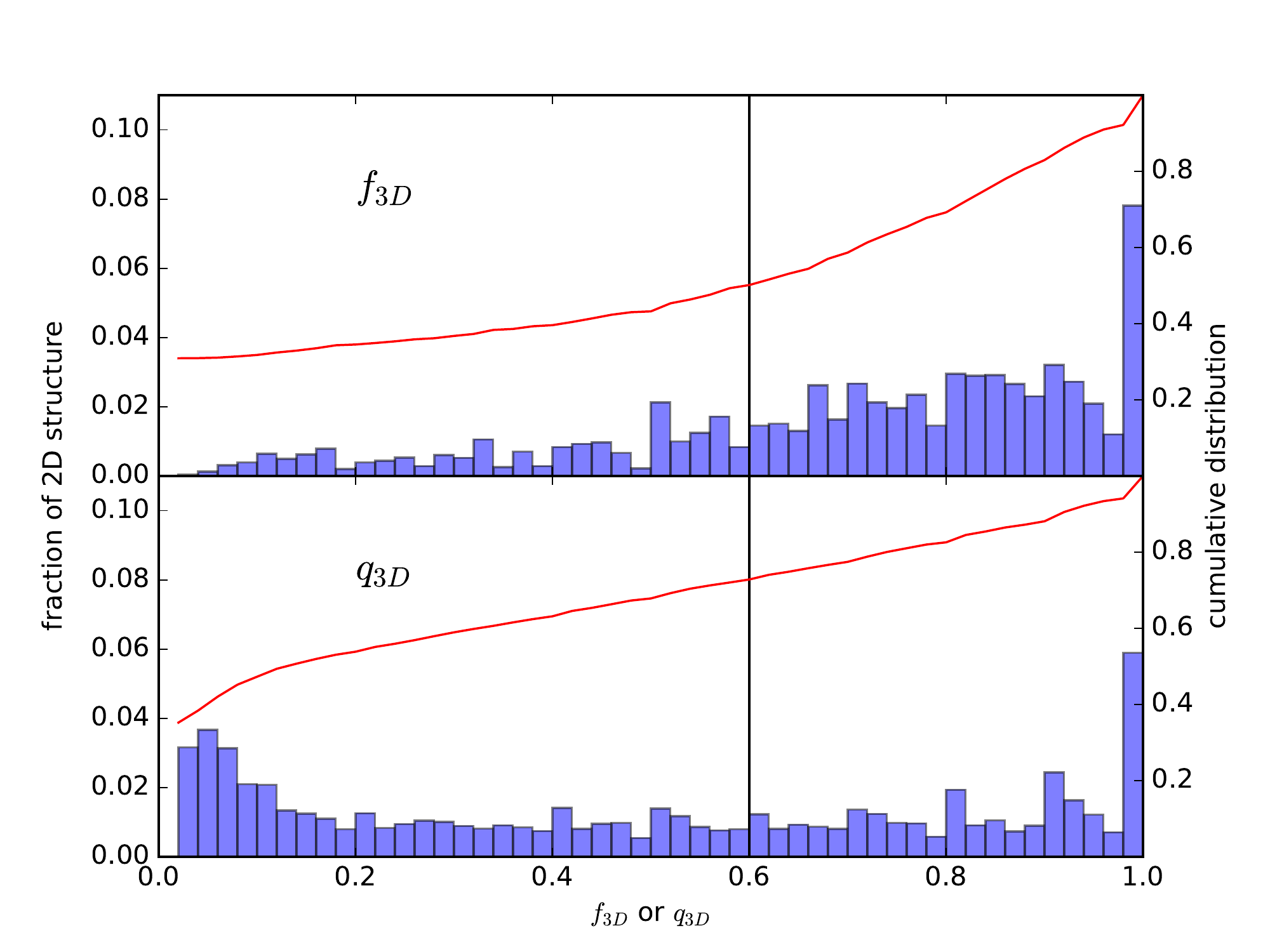}
\caption{ 
The upper panel shows the distribution of $f_{3D}$, 
the largest fraction of the total number of members of a 2D structure
that are also members of a single 3D structure.
The red solid line is the cumulative distribution function, whose value is 
shown on the right vertical axis.
The lower panel shows the distribution and the cumulative distribution function of $q_{3D}$. 
In both panels, we omit the bar corresponding to a ratio $f_{3D}$ or $q_{3D}$ smaller than 0.02 for clarity, but 
its value can be read from the cumulative profile: it is 0.41 for $f_{3D}$ and 0.45 for $q_{3D}$.
The black vertical line indicates our threshold 0.6.}

\label{f3d}
\end{figure}

The upper panel of Figure \ref{f3d} shows the differential (histogram) and cumulative (solid line) distributions of 
$f_{3D}$. The initial value of the cumulative distribution is 0.41, and we thus see that 59\% 
of the 2D structures
contain at least one member of a 3D structure. The vertical line shows the value $f_{3D}=0.6$ and 
crosses the cumulative distribution at the value 0.68, implying that 32\% of the 2D structures have $f_{3D}>0.6$. 
The median of $f_{3D}$ is 0.38. When $f_{3D}>0.6$ we assume that a 2D structure
is successfully associated to a 3D structure. We thus 
define the 2D structures with $f_{3D}>0.6$  {\it successful} 2D structures.
It can happen that different 2D structures contain members of the same 3D structure.
This event occurs for 4\% of the 2D structures.
In these cases, we take the 2D structure containing the
largest number of the 3D structure members as the possible match
to the 3D structure. 

We also compute
\begin{equation}
 q_{3D} = \frac{n (mem_{3D} \in mem_{2D}) }{n_{3D}}
\label{eq:q3D}
\end{equation}
to quantify to what extent a 3D structure is included in a 2D structure: in fact, $q_{3D}$ is the ratio between the largest number of particles, among the $n_{2D}$ members of the 2D structure, that
are also members of a single 3D structure with $n_{3D}$ members, and  $n_{3D}$ itself.
Therefore, unlike $f_{3D}$, $q_{3D}$ keeps the information on those 3D structures that are mostly or 
fully contained in a 2D structure, even if the 2D structure is not successful.
These 2D structures might 
contain too many interlopers or might be too composite to be considered successful 2D structures;
however, they still contain a substantial fraction of a 3D structure and they have thus succeeded in locating 
its presence.
The lower panel of Figure \ref{f3d} shows the differential (histogram) and
cumulative (solid line) distributions of $q_{3D}$: by looking at the values of the cumulative distribution
indicated on the right vertical axis, we see that
5.9\% of the 2D structures contain complete 3D structures;
11.9\% of the 2D structures contain more than 90\% of the members  of a single 3D structure, and
36.0\% of the 2D structures have $q_{3D}$ larger than 0.6. 
According to equation (\ref{eq:q3D}), when $q_{3D}>0.6$, more than 60\% of the members of the 3D
structure are included in the 2D structure.

According to our definitions, a one-to-one correspondence between a 2D structure and a 3D structure 
occurs when the conditions $f_{3D}>0.6$ and $q_{3D}>0.6$ are satisfied at the same time on the same
pair of 2D and 3D structures. This combined condition occurs for a 52\% and 61\% of the successful 
2D structures for the normal and merger cluster samples, respectively. 
The remaining fraction of successful 2D structures,
with $q_{3D}<0.6$, are almost exclusively associated to the cluster cores: 39\% and 25\% 
for the normal and merger cluster samples, respectively. 
In principle, a one-to-one correspondence would also be guaranteed when $f_{3D}>0.5$ and $q_{3D}>0.5$.
However, we  prefer to adopt those more restrictive thresholds to suppress the effect of noise 
around the $0.5$ value of the thresholds.

To quantify the performance of our Blooming Tree Algorithm, we define the success
rate and the completeness of the 2D structure sample. 
The success rate is 
the ratio between the number of successful 2D structures and the total number
of 2D structures:
\begin{equation}
\mathrm {Success\ Rate = { No.\ of\ 2D\ structures\ with\ f_{3D}>0.6\over  Total\ no.\ of\ 2D\ structures }} \; .
\end{equation} 
To estimate the completeness, we only consider the successful 2D structures ($f_{3D}>0.6$). Each successful 2D structure
has an associated 3D structure. The completeness is the ratio between the number of these identified 3D structures
and the total number of 3D structures in the FoV: 
\begin{equation}
\mathrm {Completeness = { No.\ of\ identified\ 3D\ structures \over  Total\ no.\ of\ 3D\ structures }} \; .
\end{equation} 

\begin{figure}[htbp]
\includegraphics[width=0.45\textwidth]{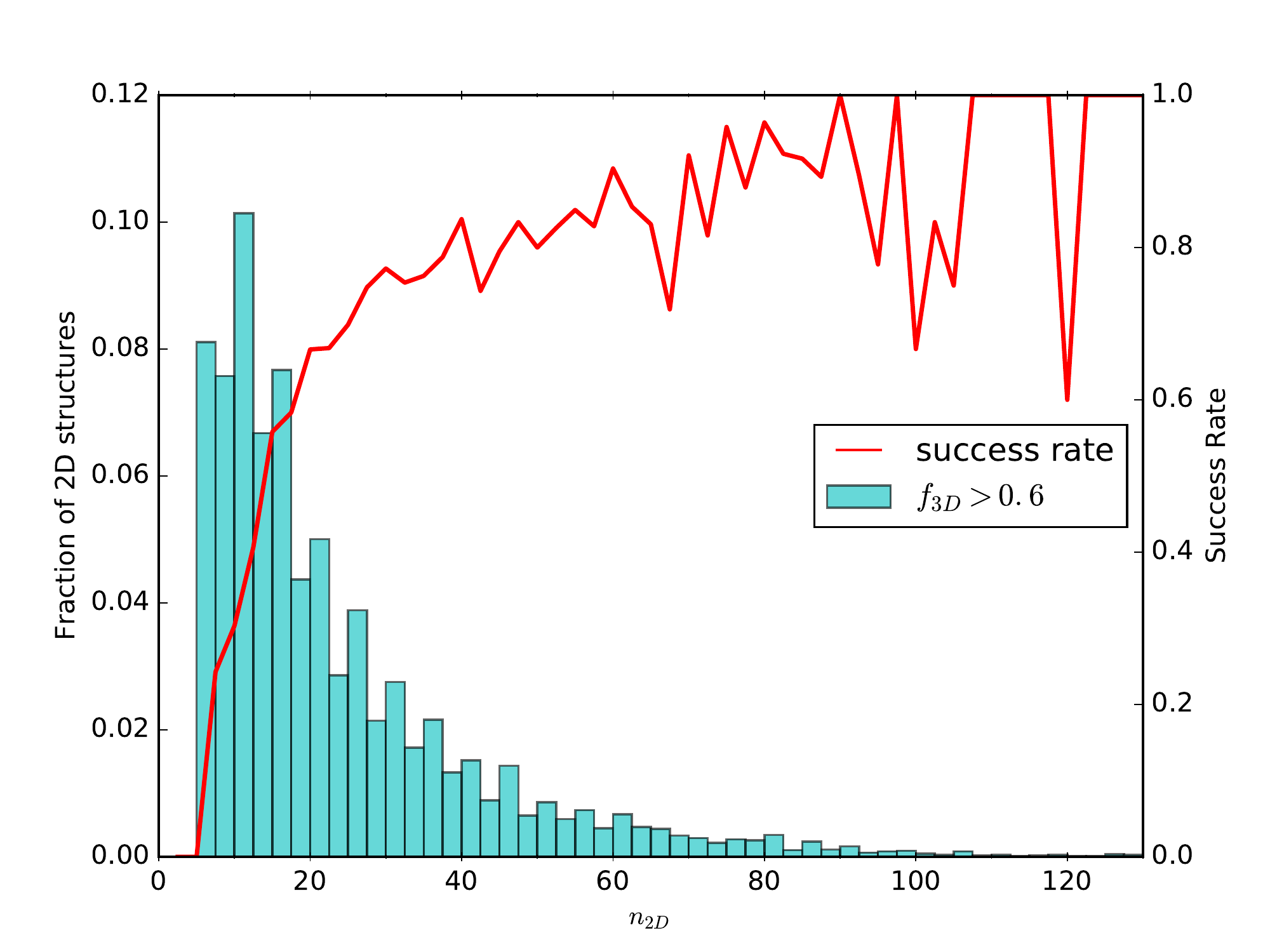}
\caption{The cyan solid histogram shows the distribution of $n_{2D}$ of the successful 2D structures, those with $f_{3D}>0.6$. 
The solid red line shows the success rate, the ratio between the number of successful 2D structures and
the total number of 2D structures as a function of $n_{2D}$ of the 2D structures.
The 15 2D structures with more than 130 members, that represent 0.2\% of the total sample}, are not shown for clarity.
\label{numhst}
\end{figure}

Figure \ref{numhst} shows the success rate as a function of $n_{2D}$, the number of members of the 2D structures
(solid line), and the distributions of $n_{2D}$ (histogram). 
The success rate, namely the probability that a 2D structure identifies a 3D structure,
clearly is proportional to $n_{2D}$ in the range $n_{2D}\lesssim 80$. The success rate flattens out, on average, at larger $n_{2D}$. 
We remove those 2D structures that are unlikely to correspond to 3D structures by setting 
a lower limit to $n_{2D}$.
We adopt the threshold $n_{2D}=10$: according to Figure \ref{numhst}, for $n_{2D}\ge 10$, 
the success rate is always larger than 30\%.

\begin{figure}
\includegraphics[width=0.45\textwidth]{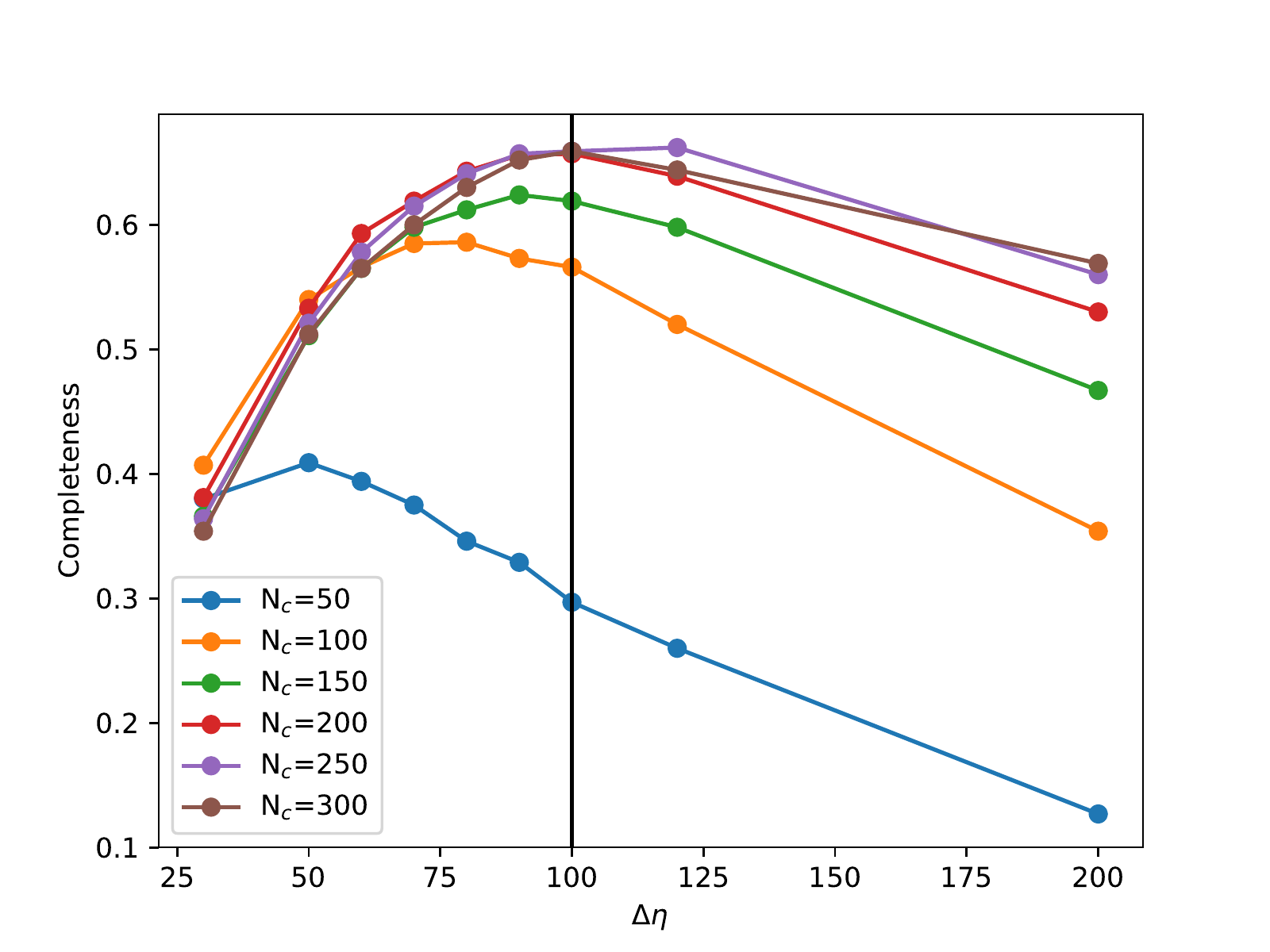}\\
\includegraphics[width=0.45\textwidth]{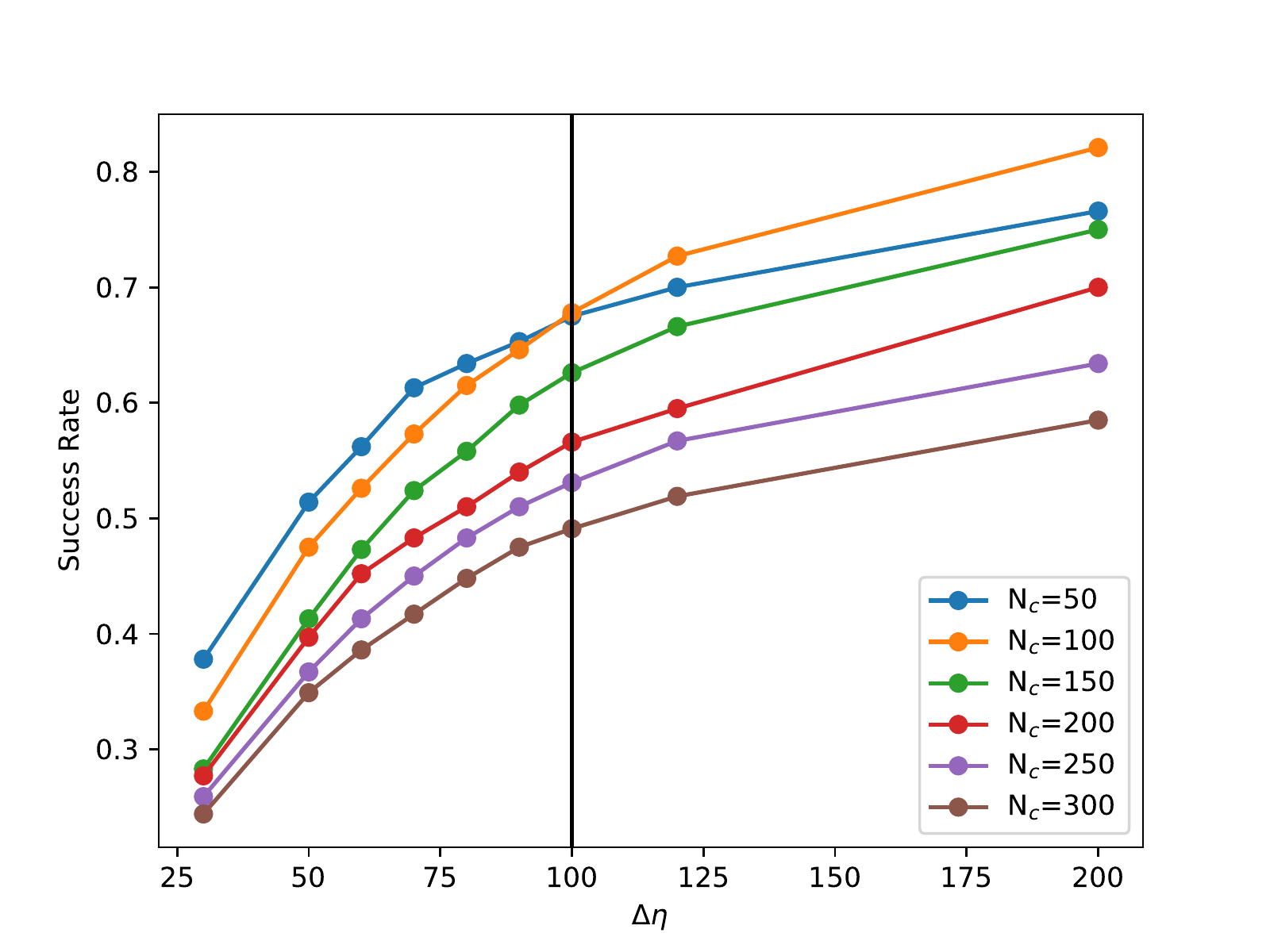}\\
\caption{The completeness and success rate as a function of the threshold $\Delta \eta$. 
}
\label{fig:eta}
\end{figure}

With the definitions of success rate and completeness at hand, we can now show how these
two quantities vary with the value of the threshold adopted for $\Delta \eta$ (Sect. \ref{sec:branch}). By increasing the threshold $\Delta \eta$, the Blooming Tree Algorithm identifies a decreasing number of 2D structures whose probability of being successful 2D structures increases. At the same time, the number of members of the 2D structures $n_{2D}$ decreases. The combination of these two effects makes the completeness peak at $\Delta \eta\sim 100$ for $N_c>150$, as shown in the top panel of Figure \ref{fig:eta}. On the contrary, the number of 2D structures decreases with $\Delta \eta$ more rapidly than the number of successful 2D structures; it follows that the success rate monotonically increases with $\Delta \eta$, as shown in the bottom panel of Figure \ref{fig:eta}. The two panels show that the successful rate increases at the expenses of the completeness. We thus adopt the threshold $\Delta \eta =100$, that maximizes the completeness for $N_c>150$. The Blooming Tree Algorithm appears to be robust against the value of this threshold: we have tested that the results presented in this work for $\Delta \eta =100$ remain substantially unaffected by adopting $\Delta \eta$ in the range $(50,120)$, where $50$ and $120$ maximize  the completeness for $N_c=50$ and $N_c=300$, respectively.  

\subsection{2D vs. 3D Structures}
\label{sec:statcomp}

We now illustrate the performance of our Blooming Tree Algorithm in identifying the 3D structures
present in the FoV. 

Table \ref{table:resu} lists the success rate and the completeness for our two 
cluster samples for different $N_c$, the number of galaxies within a sphere 
of 6~$h^{-1}$~Mpc from the cluster center; $N_{2D}$ is the average number of structures in the FoV's in addition 
to the cluster.
For the completeness, we distinguish between substructures, cores, and surrounding groups.
For a typical density of the redshift survey $N_c=200$ \citep[e.g.,][]{Rines2006, Rines2013}, Table \ref{table:resu} shows, for example, that the Blooming Tree Algorithm recovers $ 79.6$\% of the real substructures and $ 59.8$\% of the surrounding groups.

\begin{figure}
\includegraphics[width=0.45\textwidth]{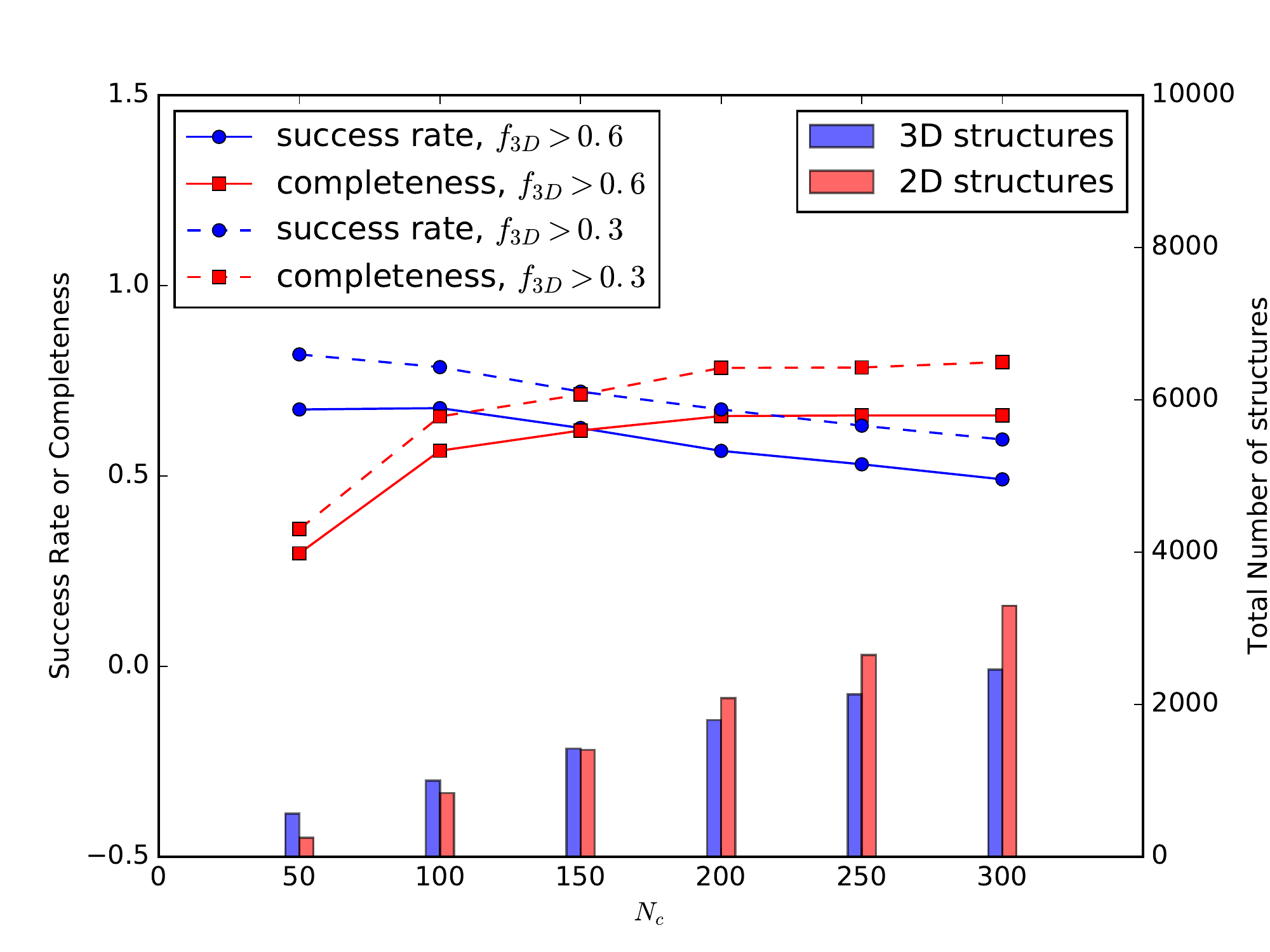}
\caption{
The success rate (blue dots) and the completeness (red squares) against $N_c$. 
The blue lines show the success rate for the 2D structures with $f_{3D}>0.6$ (solid line) and with $f_{3D}>0.3$ (dashed line).
The red lines show completeness for the 3D structures associated to the 2D structures with $f_{3D}>0.6$ (solid line) and with $f_{3D}>0.3$ (dashed line).
The blue (red) bars at the bottom show the total numbers of  
3D (2D) structures. The larger number of 2D structures indicates that
some 2D structures are spurious, namely they do not correspond to any 3D structure: the relative number of these 
spurious 2D structures is smallest when $N_c=150$.  
}
\label{fig:sum}
\end{figure}

Figure \ref{fig:sum} shows the success rate against $N_c$ for successful 2D structures
($f_{3D}>0.6$) and for 2D structures with $f_{3D}>0.3$. In this latter case, the success rate is the
ratio between the number of 2D structures with $f_{3D}>0.3$ and the total number of 2D structures. 
Figure \ref{fig:sum} also shows the completeness for 3D structures associated to 2D structures with $f_{3D}>0.6$ or $f_{3D}>0.3$. 

The success rate decreases in dense FoV's (larger  $N_c$) because of the increasing number of chance alignments. 
On the contrary, the completeness basically is independent of $N_c$, 
except for the poorest fields with $N_c = 50$, where the smaller number of galaxies in the FoV reduces the
probability of identifying the structures.
These results show that structures are satisfactorily identified in dense fields with roughly 100-150 galaxies 
within 6 $h^{-1}$~Mpc and within 4000 km~s$^{-1}$ from the cluster center: somewhat counterintuitively,
increasing the number of galaxies does not increase
the completeness and might actually decrease the success rate.

This result cannot be considered a shortcoming of the Blooming Tree Algorithm {\it tout court}: 
gravity is an infinite range force and the definition of the borders of a 3D structure is debatable. 
The trend of the success rate with $N_c$ also indicates that, in the presence of dense redshift surveys, 
the Blooming Tree Algorithm is more generous than the 3D structure identification algorithm in finding 
structures and assigning members to them. However, the Blooming Tree Algorithm also returns a completeness 
which is independent of $N_c$, when $N_c>100$, demonstrating the quite relevant ability of the Blooming Tree Algorithm 
to recover the 3D structures independently of the redshift survey density. 

\begin{table}
\caption{Success rate and completeness ($\Delta \eta$=100).}
\centering
\begin{tabular}{|c|c|c|c|c|c|c|c|}
\hline
\multirow{2}{*}{$N_{c}$} & \multirow{2}{*}{cluster} & \multirow{2}{*}{$N_{2D}$} & Success & \multicolumn{4}{|c|}{Completeness {\small (\%)}} \\
\cline{5-8}
   & &  & rate {\small (\%)} & tot & core & subs & groups  \\
\hline
\multirow{3}{*}{50}
 & normal  & 0.6 & 75.0 & 27.6 & 28.7 & 29.1 & 25.5 \\
 & merging  & 1.0 & 63.1 & 31.3 & 21.5 & 25.9 & 40.9\\
 \cline{2-8}
 & total & 0.8 & 67.5 & 29.7 & 25.2 & 27.2 & 33.6 \\
 \hline
 \multirow{3}{*}{100}
 & normal  & 2.3 & 68.8 & 54.3 & 80.7 & 76.0 & 40.7 \\
 & merging  & 3.3 & 67.1 & 58.5 & 66.0 & 59.8 & 57.0 \\
 \cline{2-8}
 & total & 2.8 & 67.8 & 56.6 & 73.3 & 65.6 & 48.8 \\
 \hline
 \multirow{3}{*}{150}
 & normal  & 3.5 & 65.2 & 53.9 & 92.0 & 77.7 & 43.6 \\
 & merging  & 5.8 & 61.0 & 68.5 & 90.0 & 77.7 & 61.9 \\
 \cline{2-8}
 & total & 4.7 & 62.6 & 61.9 & 91.0 & 77.7 & 52.8 \\
 \hline
 \multirow{3}{*}{200}
 & normal  & 5.5 & 57.6 & 58.8 & 96.7 & 79.9 & 52.0 \\
 & merging  & 8.4 & 56.0 & 71.4 & 90.7 & 79.4 & 67.2 \\
 \cline{2-8}
 & total & 7.0 & 56.6& 65.7 & 93.7 & 79.6 & 59.8 \\
 \hline
 \multirow{3}{*}{250}
 & normal  & 7.1 & 54.2 & 58.8 & 98.7 & 85.8 & 51.3 \\
 & merging  & 10.6 & 52.3 & 72.0 & 94.0 & 81.3 & 67.7 \\
 \cline{2-8}
 & total & 8.8 & 53.1 & 65.9 & 96.3 & 83.0 & 59.6 \\
 \hline
 \multirow{3}{*}{300}
 & normal  & 8.9 & 49.7 & 58.8 & 99.3 & 84.8 & 52.7 \\
 & merging  & 13.1 & 48.7 & 71.9 & 96.7 & 82.5 & 67.7 \\
 \cline{2-8}
 & total  & 11.0 & 49.1 & 65.9 & 98.0 & 83.3 & 60.3 \\
\hline
\end{tabular}
\label{table:resu}
\end{table}

a measure of distance between pairs of observations)

\begin{figure}
\includegraphics[width=0.45\textwidth]{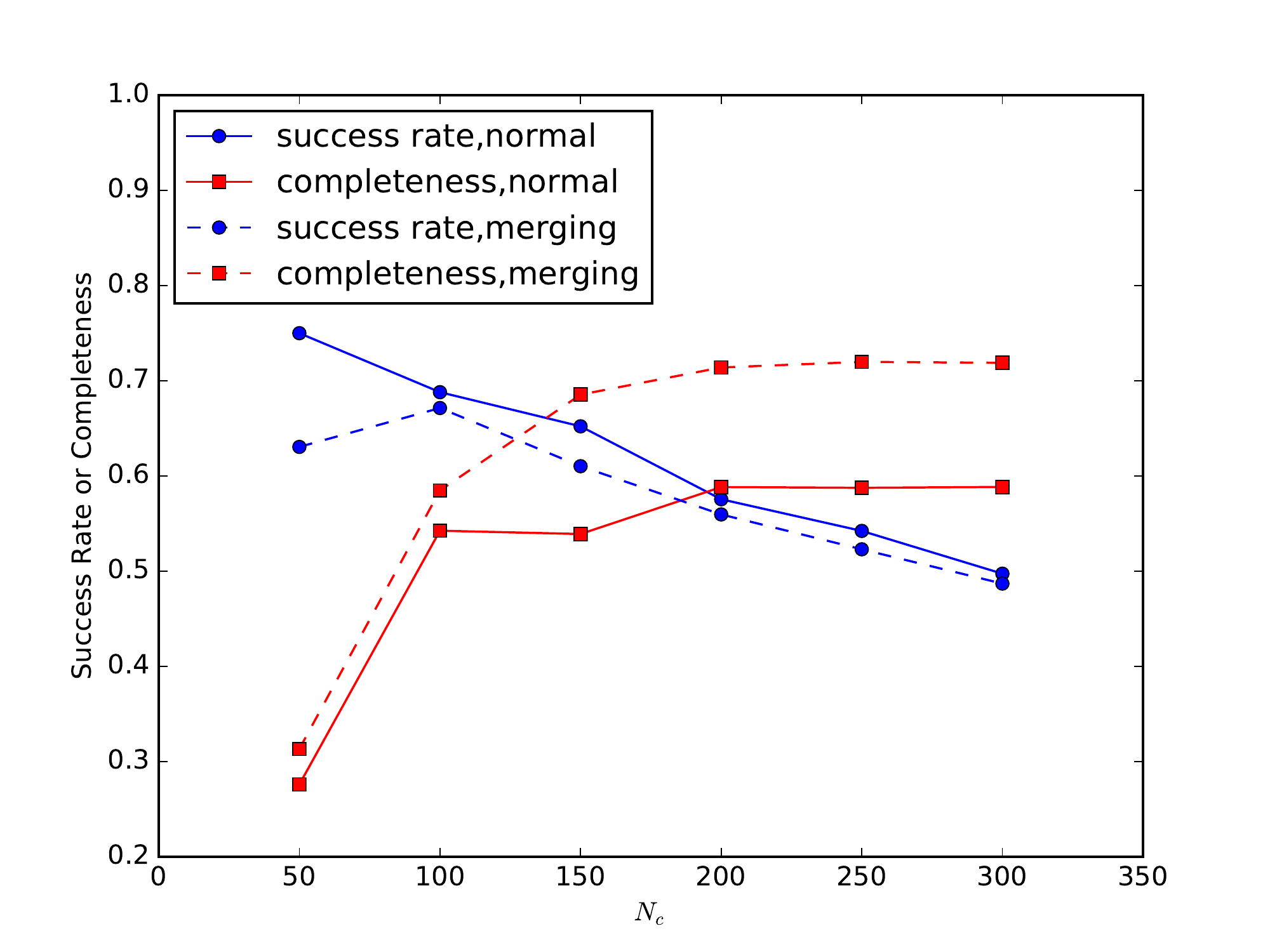}
\caption{The success rate (blue dots) and completeness (red squares) for the
merging (solid lines) and normal (dashed lines) clusters separately. } 
\label{fig:sum2}
\end{figure}

\begin{figure}
\includegraphics[width=0.45\textwidth]{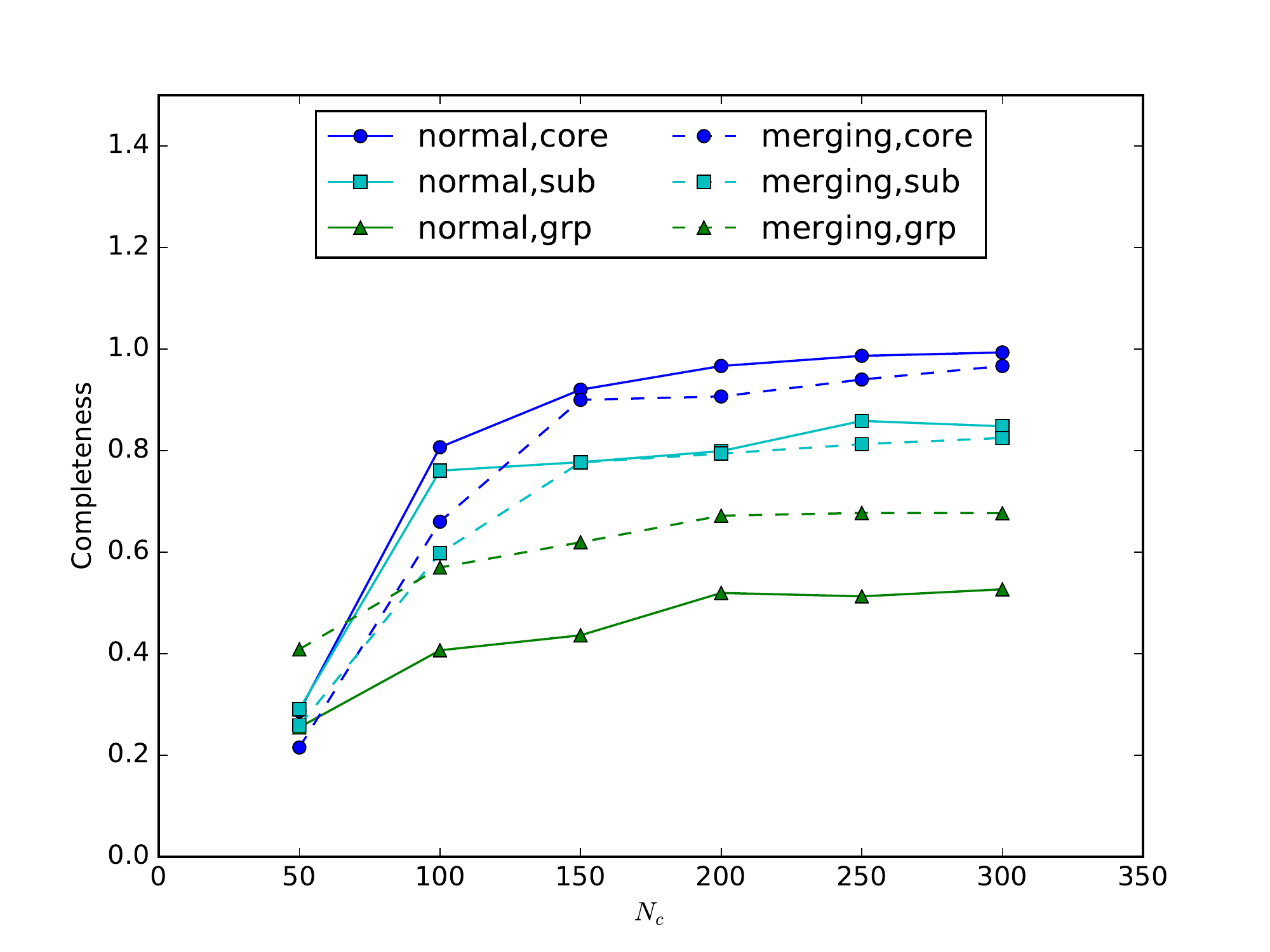}
\caption{The completeness of different types of structures: 
cluster cores (blue dots), substructures (cyan squares), 
and surrounding groups (green triangles). Solid (dashed) lines
are for normal (merging) clusters. 
}.
\label{fig:type}
\end{figure}

Figure \ref{fig:sum2} shows the success rate and the completeness 
for the merging and normal clusters separately.  
The success rate drops with increasing $N_c$ in both samples,
whereas the completeness is almost unaffected by $N_c$. In merging systems,
the galaxy distribution is more clumpy than in normal clusters; therefore, 
in merging systems the Blooming Tree Algorithm appears to be more effective and the completeness 
in this sample is systematically larger than in normal clusters. The structures
that contribute to the larger completeness in merging clusters are the surrounding groups,
as shown in Figure \ref{fig:type}.

Figure \ref{fig:type}, in fact, shows the completeness against $N_c$ of different types of structures in the two cluster samples: cores, 
substructures and surrounding groups.
In dense FoV's, the cluster cores are the easiest structures to identify, especially in normal clusters. 
Substructures require a trade off between sufficiently dense FoV's, to enable their correct identification, 
and sufficiently sparse FoV's, to minimize the interloper contamination. 
Although the surrounding groups are the most frequent structures in dense FoV's, they
show the smallest completeness ($\sim 50$\%): they are looser
sytems and are easily affected by interloper contamination.

\begin{figure}
\includegraphics[width=0.45\textwidth]{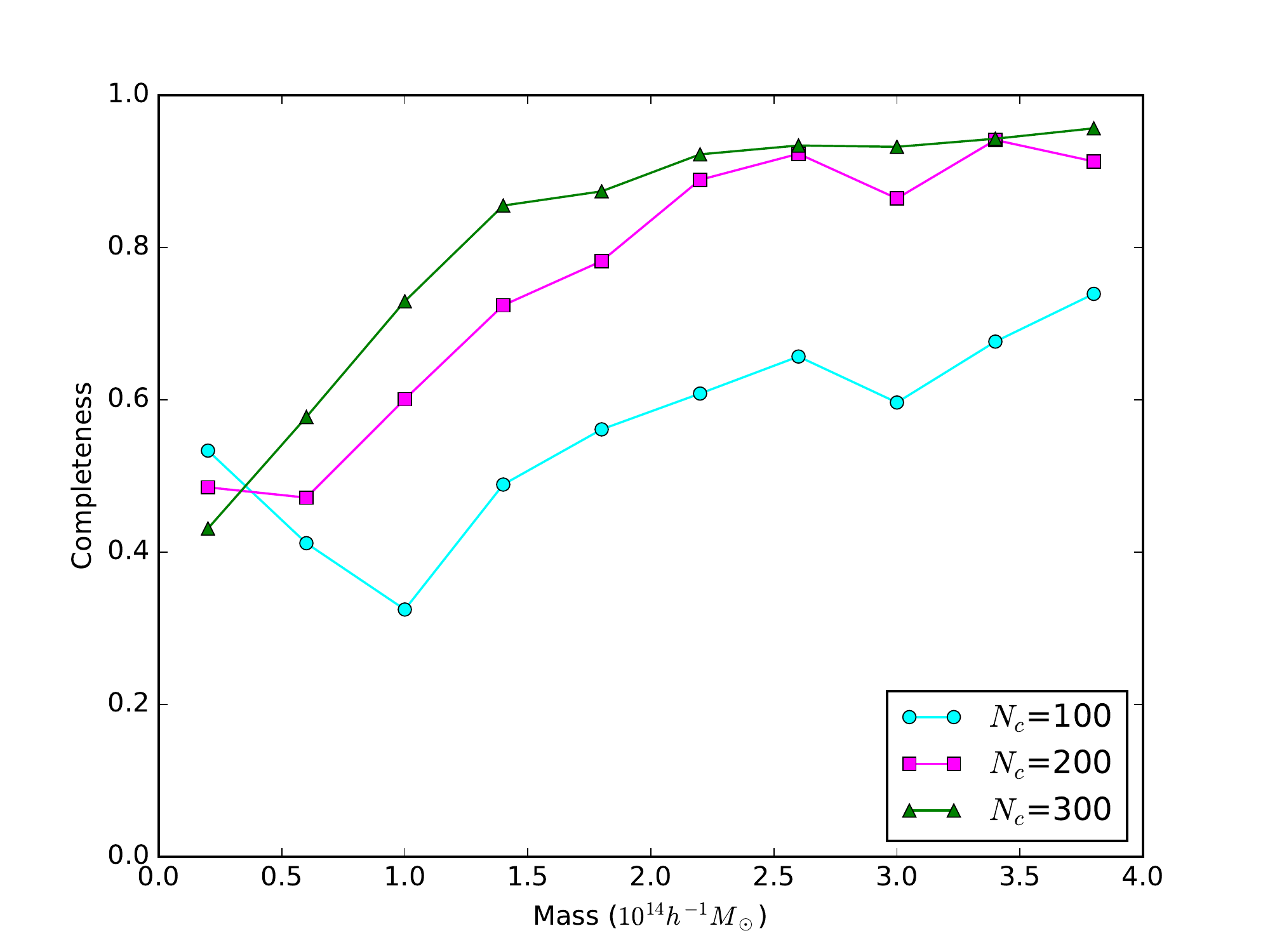}
\caption{The completeness against the 3D structure mass for different $N_c$.}
\label{fig:mhistsum}
\end{figure}

Figure \ref{fig:mhistsum} shows the completeness as a function of the mass of the 3D structure: 
massive structures with more bound members and deeper
gravitational wells are more easily detected. 
Figure \ref{fig:mhistsum} shows that, in principle, we could improve the completeness of our structure sample 
by simply dropping the less massive structures.
In the less dense FoV's ($N_c=100$ and $N_c=200$), the smallest mass bin is systematically larger than 
the second smallest mass bin: this effect is due to 
the fact that we remove all the 3D structures  
with less than 10 member galaxies from the sample of the 3D structures, 
as we mention in Sect. \ref{sec:nbody}: in fact,
in the densest FoV ($N_c = 300$), this phenomenon disappears.

\subsection{Discussion}

From Table \ref{table:resu} and Figure \ref{fig:sum}, we see that if we reduce the success threshold to  $f_{3D} > 0.3$ (Figure \ref{f3d}), 
the success rate increases from a value between 49\% and 68\%
to a value between 60\% and 82\% and the completeness increases from a value between 30\% and 66\% to a value between 36\% and 80\%. 

In addition, to be conservative, we only consider 3D structures with more than 10 particles in the fields 
(see Sect. \ref{sec:nbody}).
However, there are many 2D structures corresponding to small 3D structures with at least 5 particles; 
unfortunately, these small 3D structures are too small to increase $f_{3D}$ to $0.6$ and thus to make the 2D structure
successful. Nevertheless, if we take them into account, 
the spurious 2D structures, namely 2D structures that do not contain any 
3D structure members, will be reduced from the current 30\% to 15\%.

Finally, all these results on the success rate and completeness of our Blooming Tree Algorithm 
are based on  the {\small SUBFIND} substructure detecting algorithm  which 
is only one of the many halos finding methods adopted in $N$-body simulations.
As tested by  \citet{2013Knebe}, different halo-finding methods can give
different number of structures with a mass scatter up to 20\%.
Therefore, our strict one-to-one comparison results are certainly 
affected by the reference algorithm we adopt in the 3D simulation. Especially in the central part,
many 2D structures turn out to be failures, because the 3D halo finder algorithm associate those particles to the central core.
However, these 2D structures might well be dynamically distinct from the core.
These aspects require different analyses
that go beyond the scope of the present paper.

The above discussion clearly shows that our astrophysical problem of identifying 3D structures from the three phase-space
coordinates accessible to observations cannot be straightforwardly equated to the standard problems of clustering and classification
problems described in the classical literature of cluster analysis \citep[e.g.,][]{everitt2011,hennig2015}. In particular, 
because of the numerous possible definitions of the 3D structures and the successful 2D structures, optimization methods based on 
a minimization or maximization of a single numerical quantity cannot be adopted. Similarly, adopting classical tools, like the
receiver operating characteristic (ROC) curve and the area under curve (AUC) statistic to quantify the algorithm efficiency, is 
unfeasible, because some of the standard quantities used for their estimate, like the number of {\it true negative} elements of the data set,
cannot be defined in our problem.

\section{Comparison with the $\sigma$ plateau algorithm }
\label{sec:caustic}

We now compare our Blooming Tree Algorithm for the identification of structures
with the $\sigma$ plateau algorithm. We will first illustrate the differences
on a representative case and we will then consider the statistical 
properties of a large cluster sample.

\subsection{A Representative Case}
\label{sec:repCase}

We consider
a massive cluster with $M_{200} = 8.66 \times 10^{14} h^{-1} M_{\odot}$ ,
with 722 galaxies in the square FoV of size equal to $3 R_{200}$; 200, out of these 722 galaxies, are cluster members. 
The weak point of the $\sigma$ plateau algorithm is its difficulty to identify structures
with widely different velocity dispersions \citep{2015Yu}.
A massive cluster like the one we choose here provides structures with different
mass and size and provides thus a good test for the performance of the two algorithms.

\begin{figure}[htbp]
\includegraphics[width=.45\textwidth]{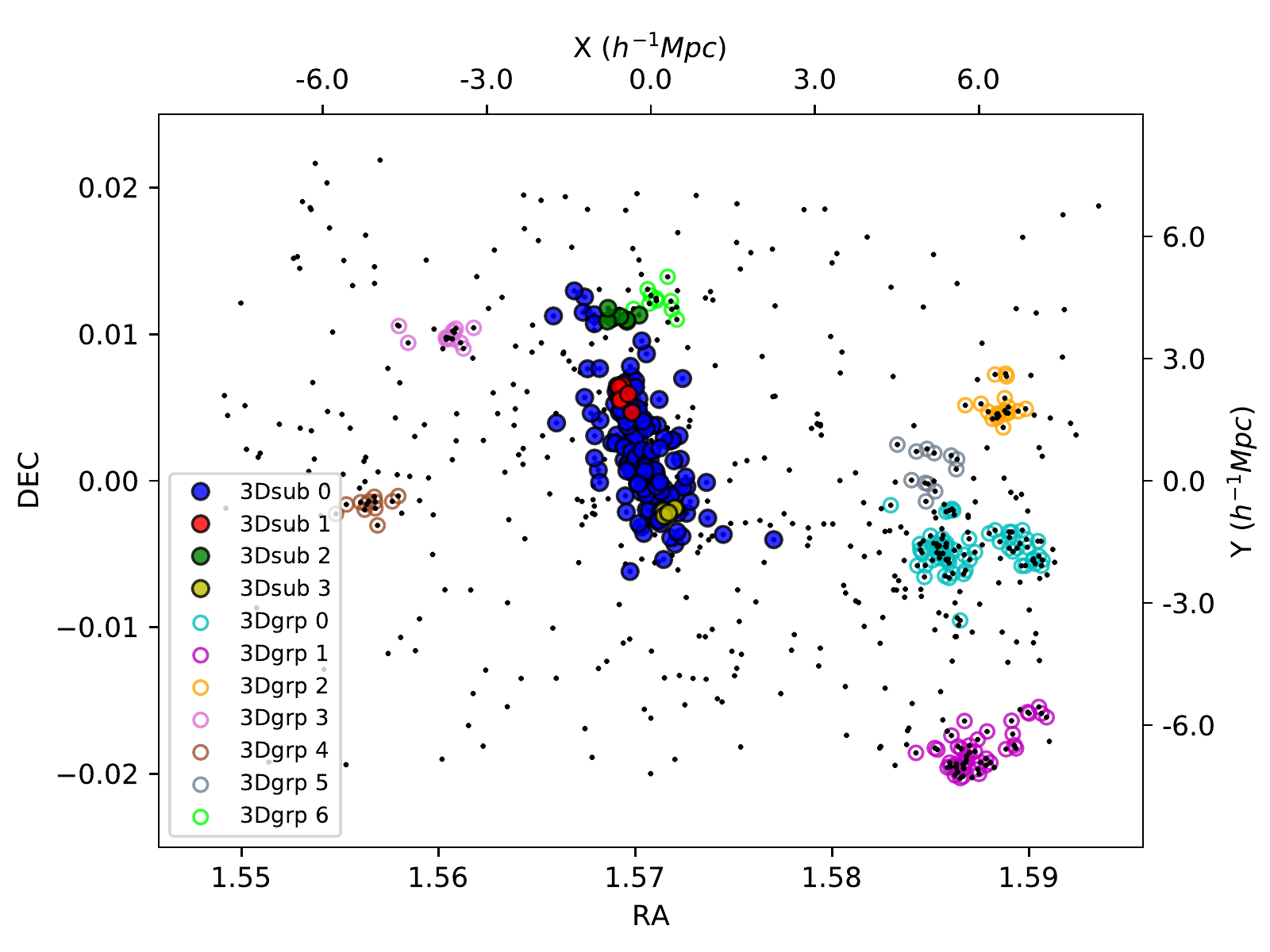}
\includegraphics[width=.45\textwidth]{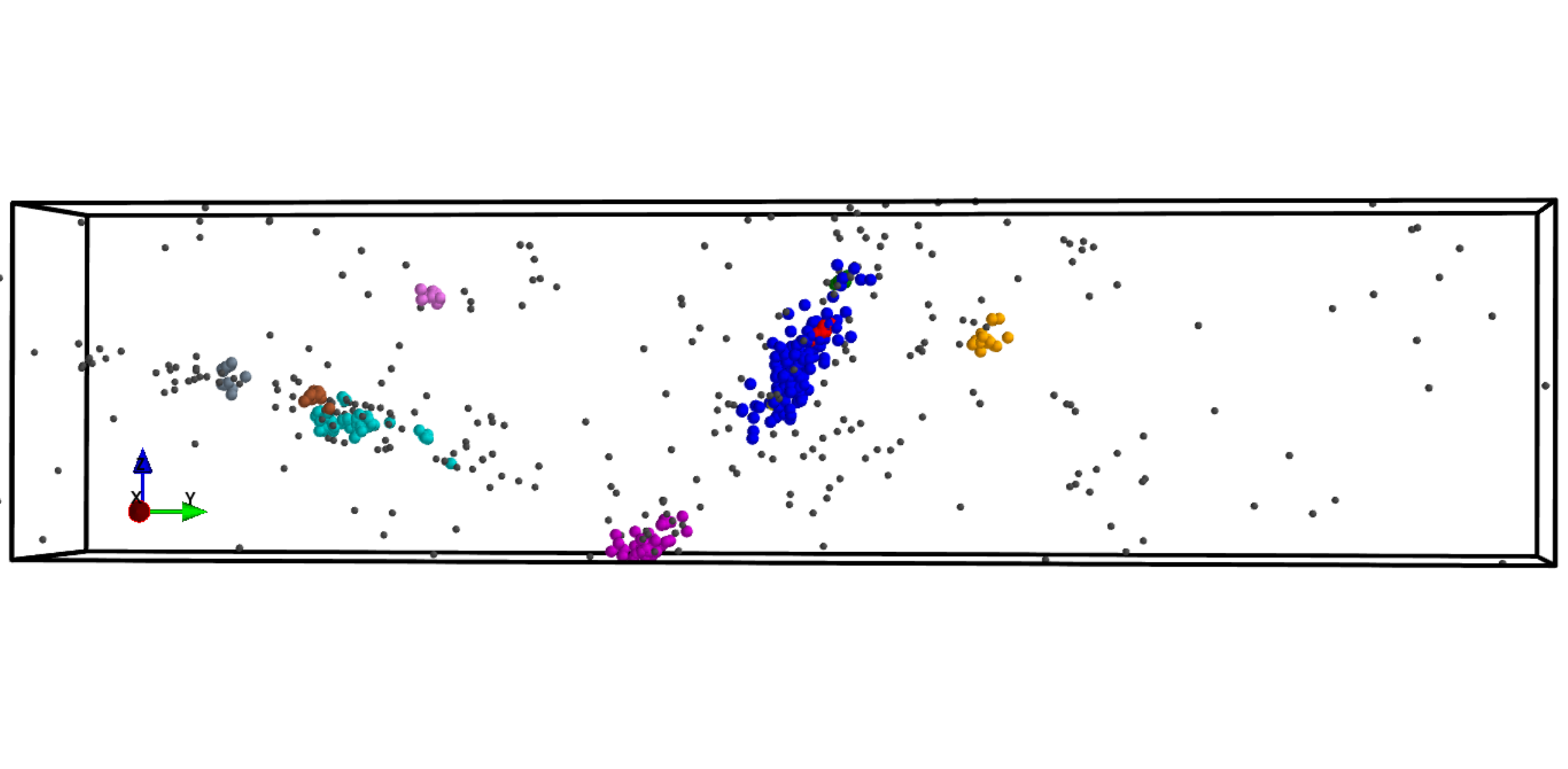}
\caption{Distribution of the 3D structures in the FoV of a massive cluster with 722 particles.
In the top panel, the projected celestial coordinates are in radiants; 
comoving coordinates in the $N$-body simulation are also shown. 
The solid dots show the members of the structures with different colors; the core particles are in blue.
The open circles show the members of the surrounding groups. 
The bottom panel provides a 3D view within a box with
dimensions $14 \times 14 \times 60 h^{-3}$~Mpc$^3$. Group 6 (open green circles) is outside this box.
} 
\label{fig:realsky2}
\end{figure}

\begin{figure*}[htbp]
\centering
\includegraphics[width=0.6\textwidth]{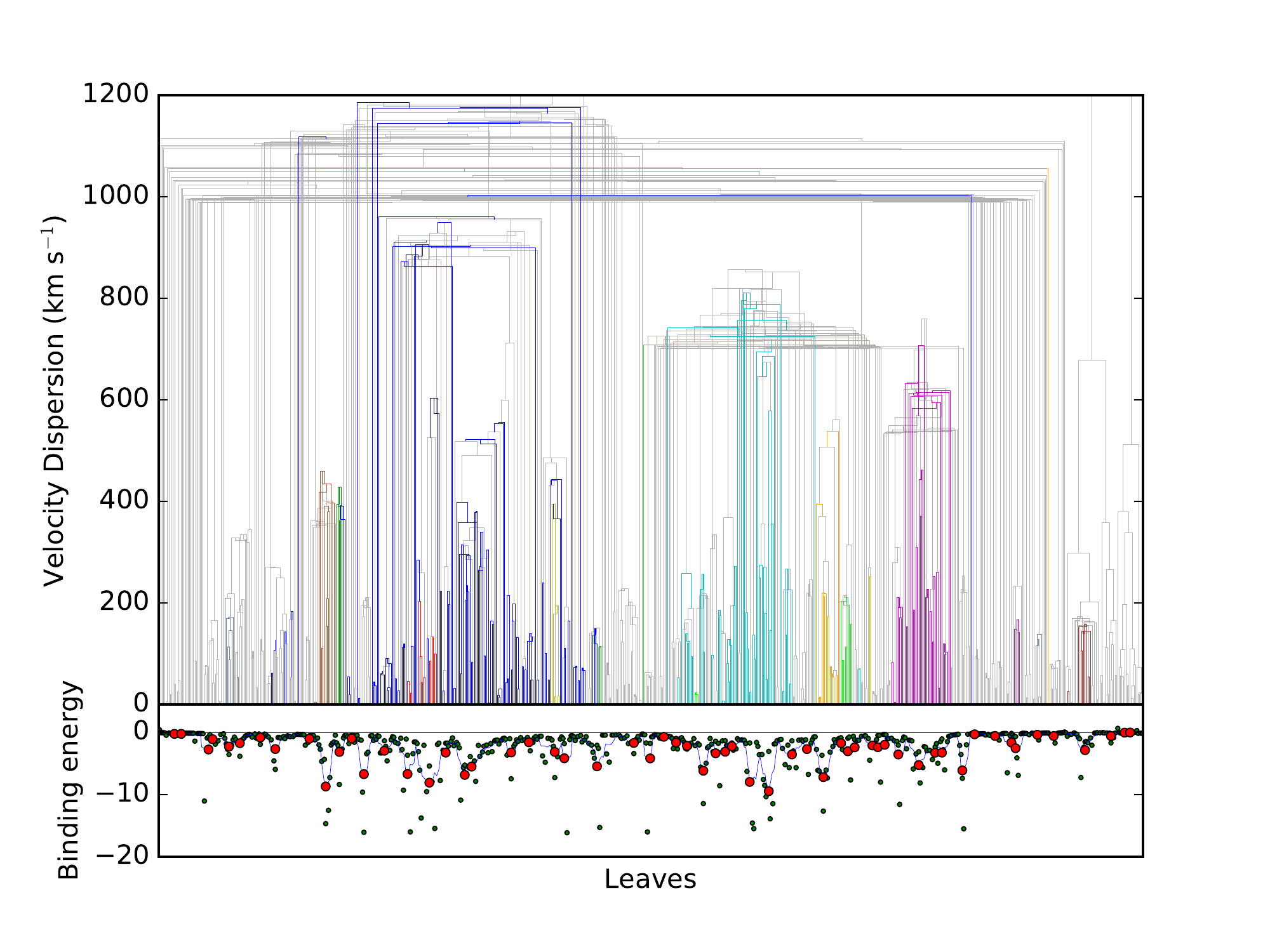}
\caption{The binary tree of the simulated massive cluster with 722 particles in the FoV
shown in Figure \ref{fig:realsky2}. 
The inset at the bottom shows the binding energy profile. 
The colors indicate the real members of the different 3D structures, 
according to Figure \ref{fig:realsky2}.}
\label{fig:dendrogram2}
\end{figure*}

\begin{figure}[htbp]
\includegraphics[width=0.48\textwidth]{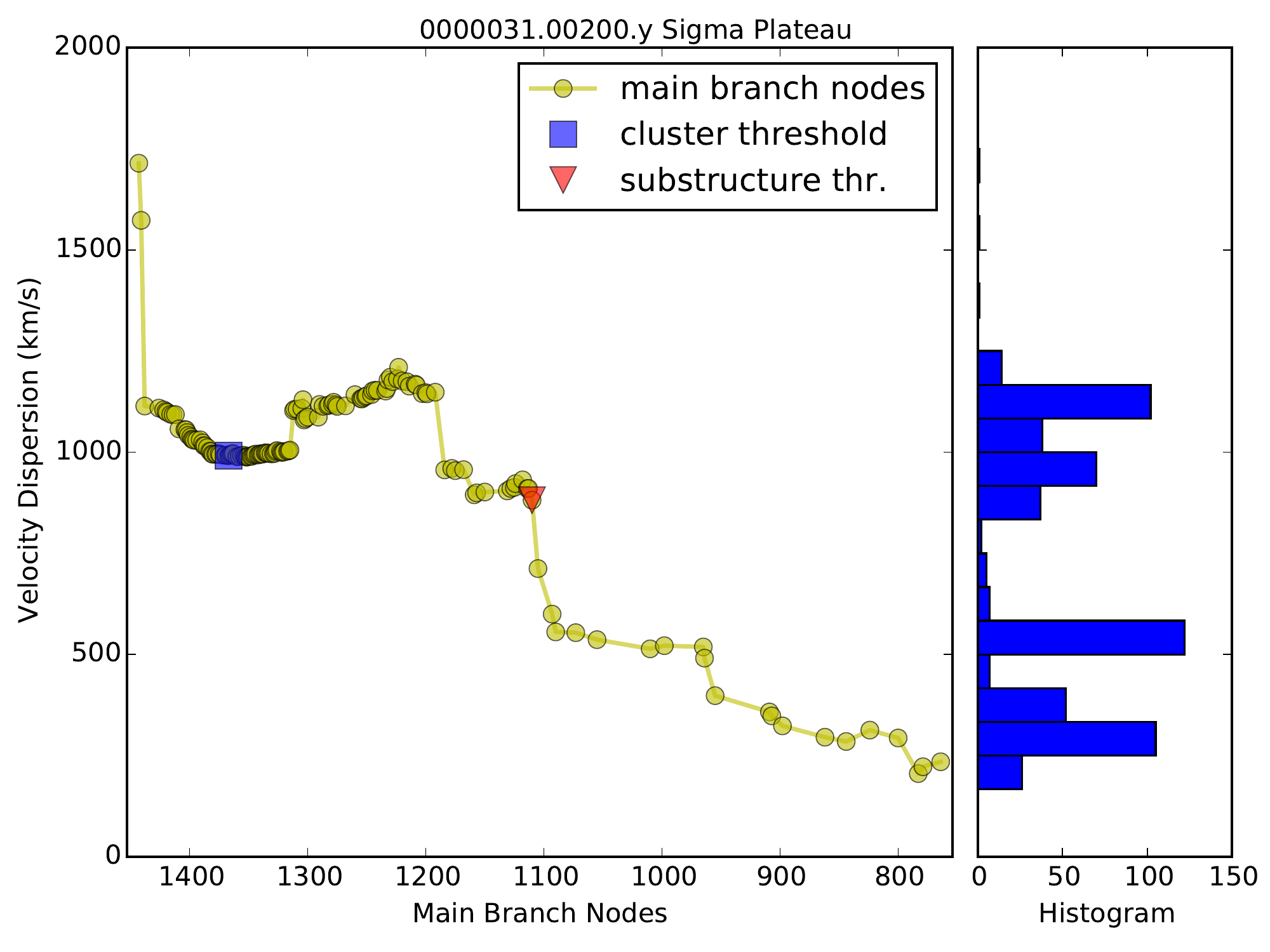}
\caption{Velocity dispersion of the leaves of each node
along the main branch of the binary tree shown in Figure \ref{fig:dendrogram2}. 
The blue square and the red triangle are the first and the second threshold, respectively. 
The curve in between is the $\sigma$ plateau, 
whose position generally corresponds to the peak of the distribution of
node numbers with similar velocity dispersion shown in the right panel. This peak does not stand out clearly here because of the 
existence of other plateaus.  }
\label{fig:plateau2}
\end{figure}

\begin{figure}[htbp]
\includegraphics[width=.45\textwidth]{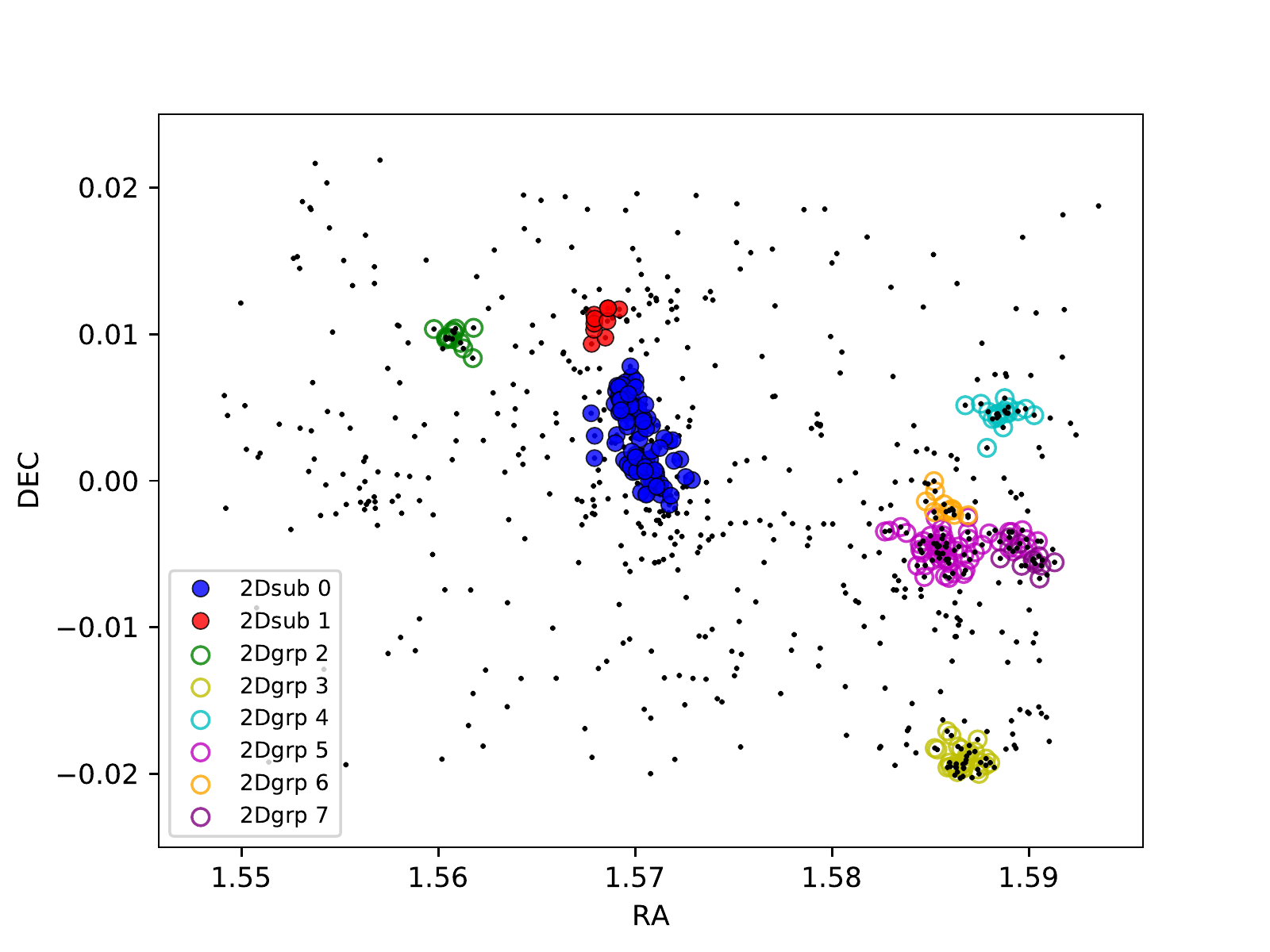}
\includegraphics[width=0.45\textwidth]{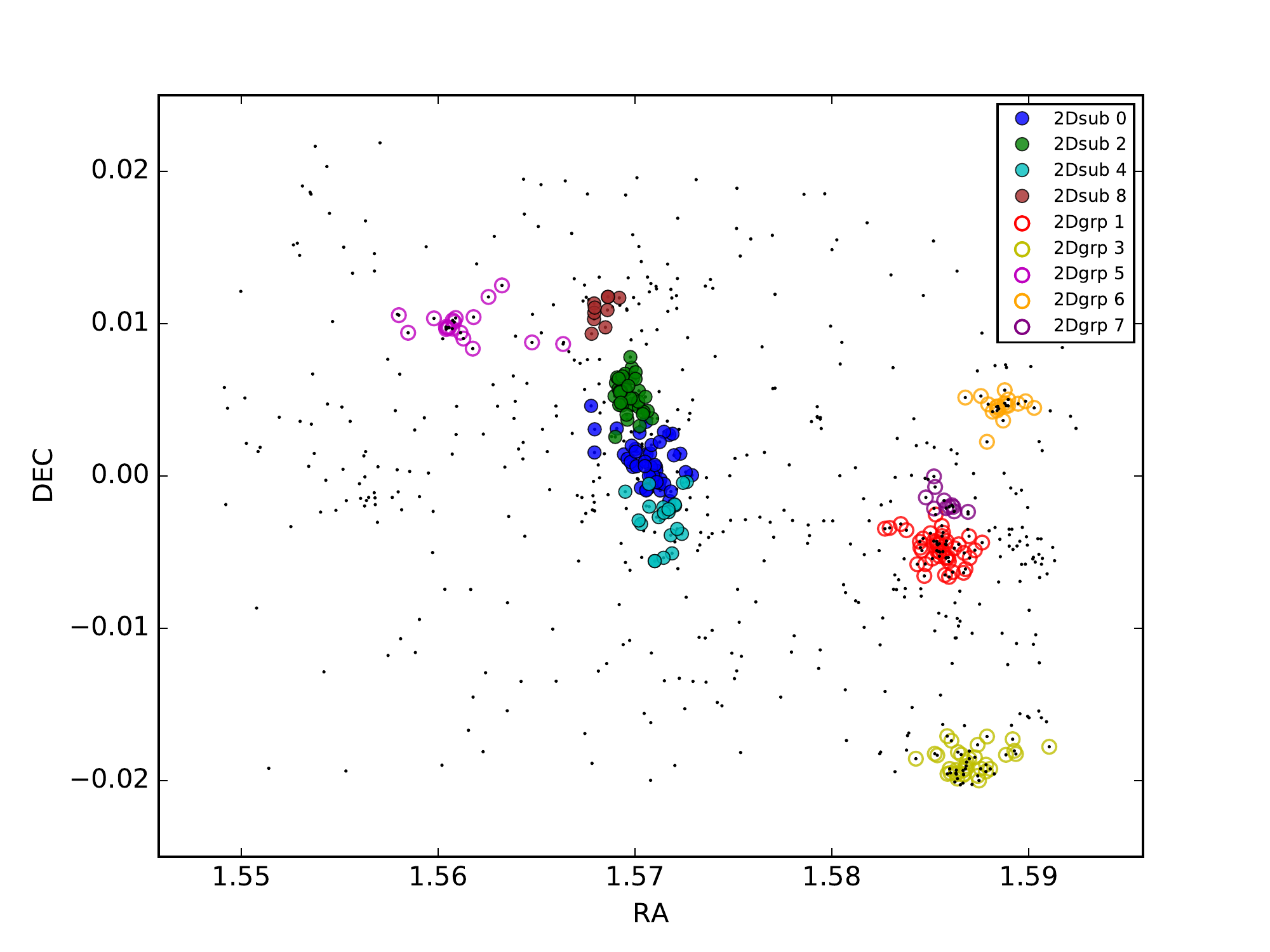}
\caption{Sky diagrams of structures identified by the $\sigma$ plateau algorithm (top panel)
and the Blooming Tree Algorithm (bottom panel).
Galaxies with the same color belong to the same 2D structure.
The galaxies in the bottom panel are colored according to colors of the binary tree in Figure \ref{fig:vtree2}. 
} 
\label{fig:result}
\end{figure}

There are four 3D substructures, including the cluster core, and seven 3D surrounding groups in the FoV, as shown in 
Figure \ref{fig:realsky2}. The distribution of these structures in the binary tree, shown 
in Figure \ref{fig:dendrogram2} with different colors, indicates  
that the member galaxies of these structures tend to appear on the
same branch of the tree; however, 
their velocity dispersions are in the range
$\sim 200-1000$ km~s$^{-1}$. As mentioned above, this large velocity dispersion range 
is the main challenge for the $\sigma$ plateau algorithm.

According to the
$\sigma$ plateau algorithm \citep{2015Yu}, by walking on the main branch of the tree, 
we determine the $\sigma$ plateau 
shown in Figure \ref{fig:plateau2}.
The large range of velocity dispersions, which is a consequence of the complex dynamics of 
clusters like this one, prevents the plateau from being flat: 
therefore, locating the second threshold in this kind of systems becomes
rather ambiguous. 

Nevertheless, the result of the $\sigma$ plateau algorithm, shown in the top panel of Figure \ref{fig:result},
are satisfying. 
The algorithm recovers two substructures out of four and six surrounding groups out of seven. 
The first 2D substructure (2Dsub 0) is the cluster core, whereas the second 2D substructure (2Dsub 1) includes
the 3D substructure 3Dsub 2; however, it does not satisfy the $f_{3D} > 0.6$ 
criterion and cannot be considered a successful 2D substructure.
The remaining 3D substructures, 3Dsub 1 and 3Dsub 3, remain unidentified, because they are 
located in the cluster center and the method is unable to separate them from  the core. 
This result is a consequence of the fact that the $\sigma$ plateau algorithm cuts the binary tree 
with a single velocity dispersion threshold, and can thus
only identify structures whose velocity dispersion is close to this threshold:
if a substructure has a small number of members, it  
can not generate a significant plateau, it is not recognized
by the algorithm as a distinct structure and  it will thus be included in a larger system.

\begin{figure*}[htbp]
\centering
\includegraphics[width=0.8\textwidth]{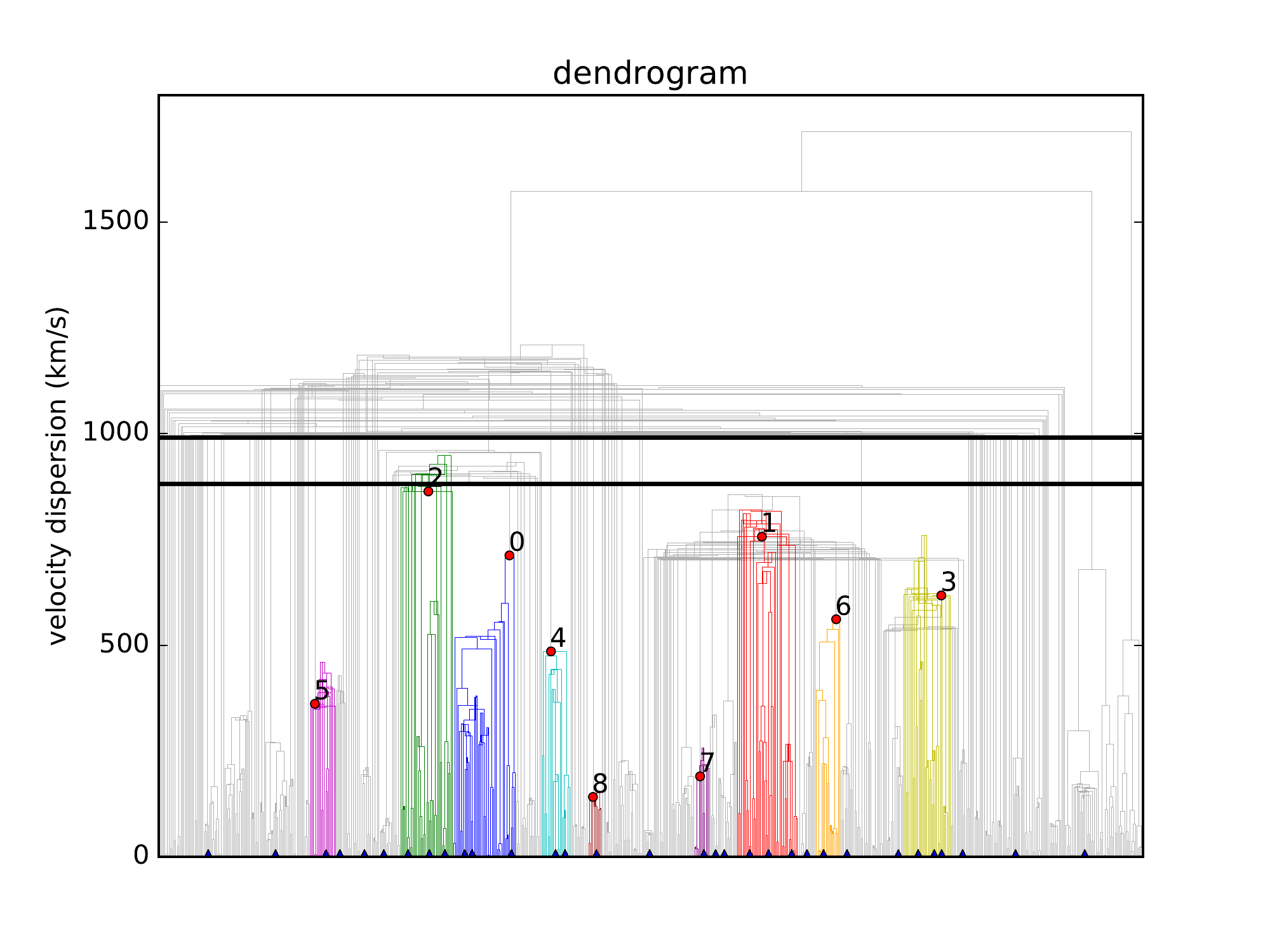}
\caption{The dendrogram with the node velocity dispersion on the vertical axis. 
The blue triangles at the bottom show the buds, the binding energy minima that identify the branches that need to be searched. 
The red dots show the blooms, all the key nodes found with the $\Delta \eta$ threshold, that identify the
2D structures.
The stems of their member leaves are shown with different colors. The structure indexes shown on the plot are sorted according to decreasing number of members.
The two horizontal black solid lines are the thresholds
of the $\sigma$ plateau algorithm.}
\label{fig:vtree2}
\end{figure*}

Figure \ref{fig:vtree2} and the bottom panel of Figure \ref{fig:result} show the results of our 
Blooming Tree Algorithm, based on tracing all the  tree branches rather than the main branch alone. 
The two solid lines in Figure \ref{fig:vtree2} indicate the two thresholds of the $\sigma$  plateau approach.
The new algorithm is able to pick up structures with different velocity dispersions. 
It recovers all the four substructures; it also recovers five out of the seven 3D surrounding groups.

The 2D substructure 2Dsub 0, that identifies the cluster core, 
has a velocity dispersion corresponding to the second threshold 
of the $\sigma$ plateau; with the first key node below the $\sigma$ plateau, 
the algorithm can identify the 2D substructure 2Dsub 2: it includes the 3D substructure 3Dsub 1 
that was unidentified by the $\sigma$ plateau algorithm. 
Finally, the 2D substructure 2Dsub 4 includes the 3D substructure 3Dsub 3, 
and the 2D substructure 2Dsub 8 includes the 3D substructure 3Dsub 2.

All the 2D structures identified by the Blooming Tree Algorithm are located at positions 
consistent with the corresponding 3D structures. However, 
only four groups, 2Dgrp 1, 2Dgrp 3, 2Dgrp 5, and 2Dgrp 6, out of five satisfy the $f_{3D} > 0.6$ criterion. 
All the three 2D substructures other than the core (the 2D substructure 2Dsub 1)
are contaminated by core members and fail the $f_{3D} > 0.6$ criterion. 
Therefore, even when the 2D and 3D structures share the same position on the sky, 
this strict criterion on $f_{3D}$ does not allow the association of the 2D structure to the
corresponding 3D structure, and the 2D structure is classified as spurious. This example shows
that the $f_{3D} > 0.6$ criterion guarantees a robust identification of the 3D structures
but it returns a lower limit to the performance of our algorithms for the identification
of structures. 

\subsection{Performance on Two Cluster Samples}
We analyze 
the two samples of simulated clusters used in \citet{2015Yu} with both algorithms.
The two samples are a massive sample (M15) containing 150 mock redshift surveys,
with $M_{200}$ ranging from $0.86 \times 10^{15} h^{-1} M_{\odot}$ 
to $3.4 \times 10^{15} h^{-1} M_{\odot}$; and a less
massive sample (M14) with the same number of clusters, 
with mass ranging from $0.95 \times 10^{14}h^{-1} M_{\odot}$ 
to $1.1 \times 10^{14} h^{-1} M_{\odot}$. 
Each cluster is sampled with a given number of galaxies within $3R_{200}$: $N_c=(100, 200, 300)$.

\begin{figure}
\includegraphics[width=0.45\textwidth]{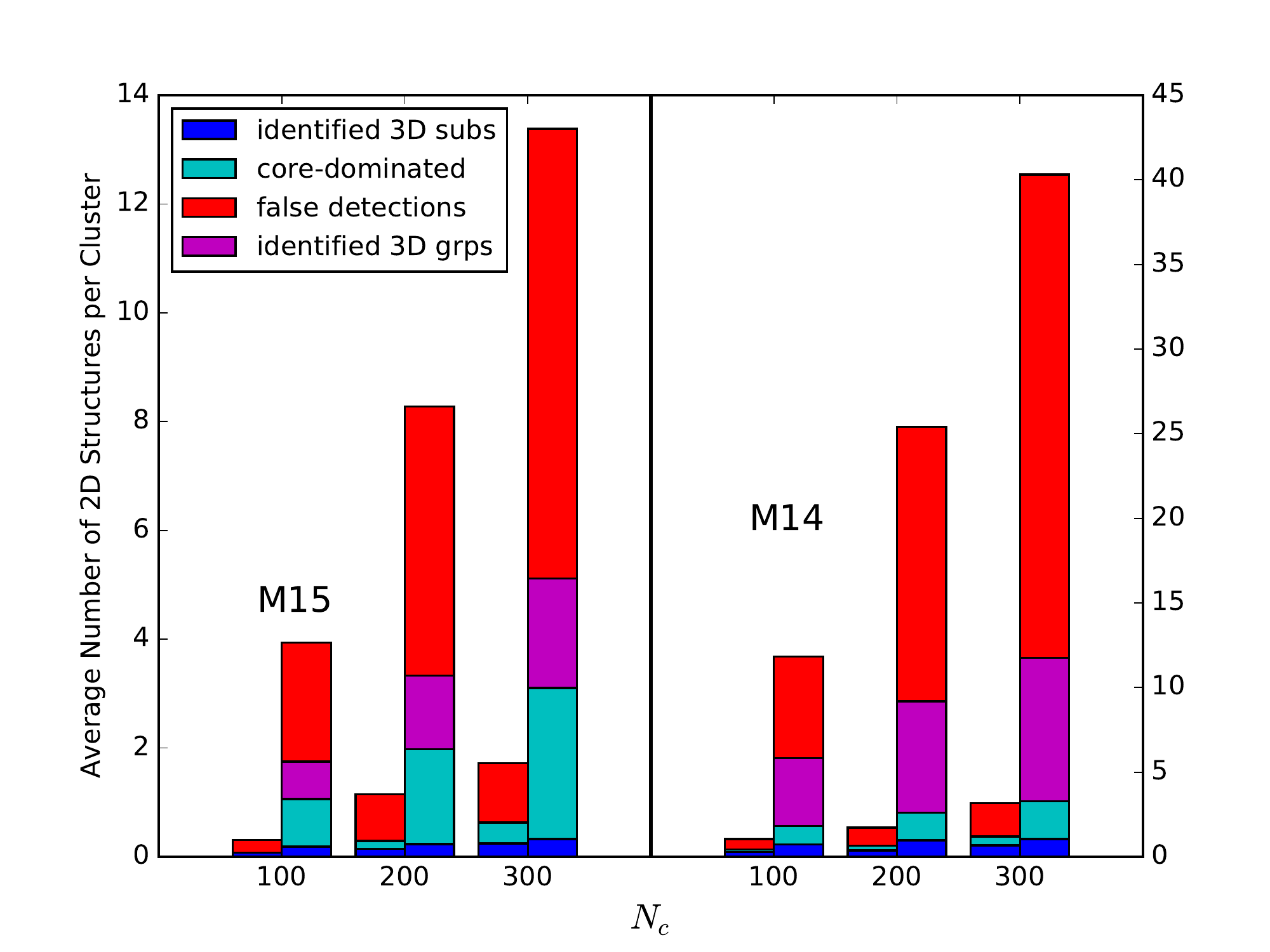}
\caption{The average number of 2D structures per cluster as a function of $N_c$: 
left (right) panels are for the M15 (M14) sample, 
left (right) bars are for the $\sigma$ plateau algorithm (Blooming Tree Algorithm). The blue
part of each bar shows the 2D substructures associated to 3D substructures ($f_{3D}>0.6$);
the cyan part shows the 2D structures associated to 3D cores ($f_{3D}>0.6$); 
the red part shows the spurious 2D structures ($f_{3D}<0.6$). The purple part shows the 2D groups
associated to 3D groups ($f_{3D}>0.6$): they are missing in the left bars, 
because the 3D groups were not included in the $\sigma$ plateau analysis of \citet{2015Yu}.}
\label{fig:comp}
\end{figure}

Figure \ref{fig:comp} shows the results of this analysis.
It shows the average number of 2D structures per cluster as a function of $N_c$.  The left 
and right bars are for the $\sigma$ plateau algorithm  and the Blooming Tree Algorithm, respectively.
The Blooming Tree Algorithm identifies five to ten times more structures than
the $\sigma$ plateau algorithm, with more structures identified in the M14 sample. In fact, 
because of our strategy for the mock catalogue creation  
(see \citealt{2015Yu}), the fields of the M14 clusters are denser than the M15 fields, and
detecting 2D structures is more likely. The two algorithms identify
a comparable number of substructures (blue sectors of the bars), whereas the Blooming Tree Algorithm
is more efficient at separating the cores into distinct structures (cyan sectors of the bars). The Blooming Tree Algorithm
also identifies the surrounding groups (purple sectors), that are not included in the $\sigma$ plateau
analysis. 

The red sectors show the false detections, the 2D structures with $f_{3D}<0.6$: 
for the Blooming Tree Algorithm they represent
$\sim 56-62\%$ of the 2D structures in the massive clusters, depending on $N_c$, and $\sim 67-77\%$ of the 2D structures
in the less massive clusters. The $\sigma$ plateau algorithm has a substantially comparable 
performance, but on a substantially smaller number of 2D structures: those fractions become $\sim 40-70\%$ and $\sim 60\%$, respectively.
These fractions would clearly decrease by adopting a weaker criterion than $f_{3D}>0.6$.
The representative case illustrated in the previous subsection \ref{sec:repCase} shows that this criterion might indeed be too strict.

The Blooming Tree Algorithm improves the completeness of the $\sigma$ plateau algorithm by 
roughly a factor of three. In the M14 sample, the  Blooming Tree Algorithm identifies 68\% of the 
cluster substructures and cores, compared to 25\% of the $\sigma$ plateau. In the M15 sample, the improvement
is even more dramatic, with a completeness of 77\% for the  Blooming Tree Algorithm and 5-20\%
for the $\sigma$ plateau.

Overall, the M15 sample has larger success rates than M14, because 
in massive clusters cores and substructures are more massive and are thus easier to identify.
In addition, 
the fields of the M14 clusters are denser than the M15 fields, and 
the probability of detecting 2D structures that do not correspond to any 3D substructure increases.
 
\begin{figure}
\includegraphics[width=0.45\textwidth]{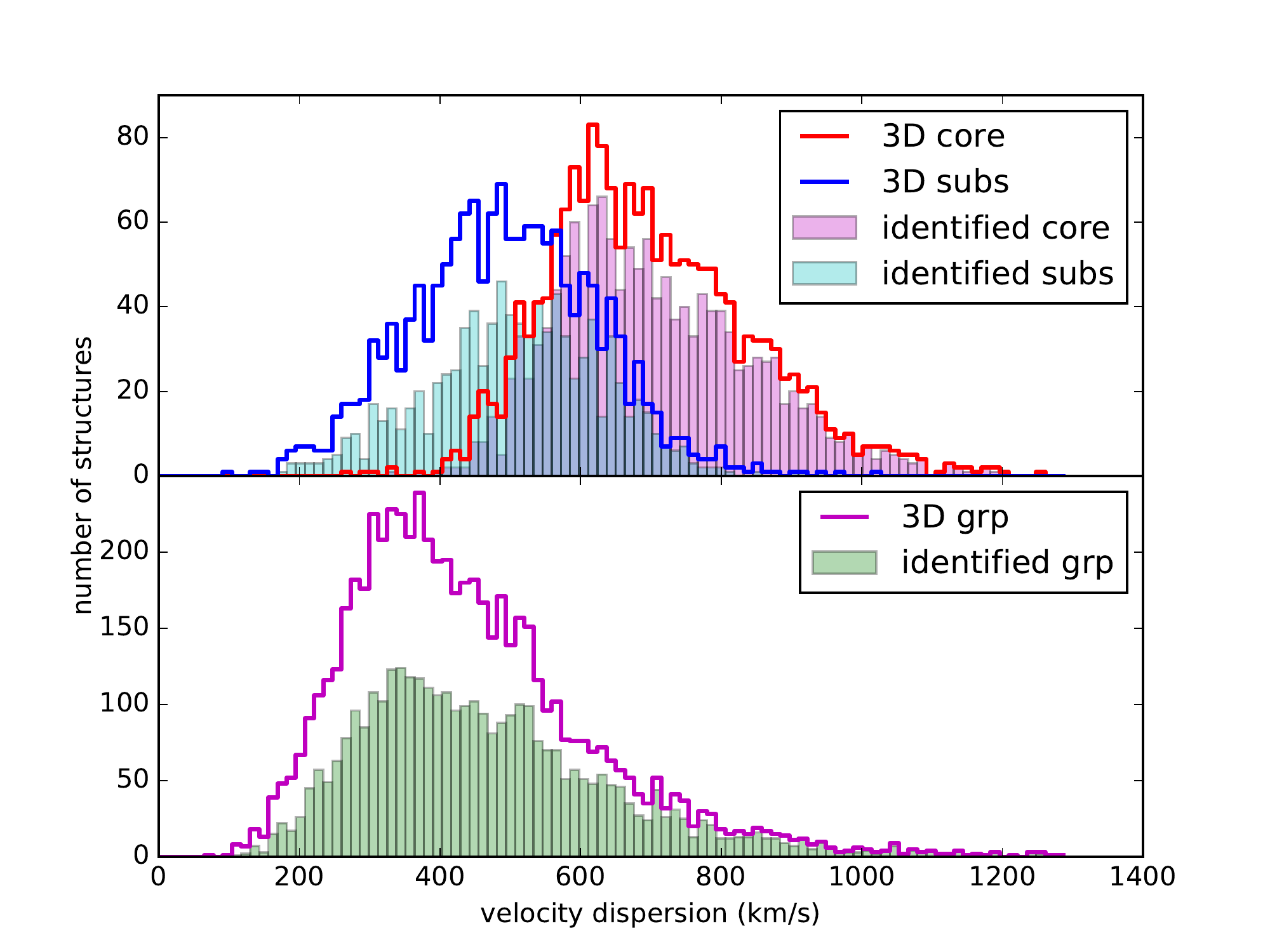}
\caption{The distributions of the velocity dispersions of different types of structures in the combined sample of 50 normal clusters and 50 merging clusters:
cores of clusters (red open histogram), substructures (blue open histogram), and surrounding groups (magenta
open histogram). 
The solid histograms are the 3D structures identified by the 
successful 2D structures with $f_{3D}>0.6$ according to the Blooming Tree Algorithm, in the 300 mock catalogues created by projecting each cluster along three orthogonal lines of sight: 
core of clusters (magenta), substructures (cyan), and surrounding groups (green).}
\label{fig:vhist}
\end{figure}

As mentioned above, the origin of the different performance between the two algorithms is 
the wide distribution of the velocity dispersions of the structures. 
Figure \ref{fig:vhist} shows the distributions of the velocity dispersions of the 
three kinds of structures in our combined sample of 50 normal clusters and 50 merging clusters: 
cores (red open histogram), substructures (blue open histogram), and surrounding 
groups (magenta open histograms).  
Unlike the $\sigma$ plateau algorithm that searches the main branch alone and only relies on the velocity
dispersion of the leaves on the main branch, the Blooming Tree Algorithm searches
all the tree branches and combines three quantities, velocity dispersion, richness, and size of 
each node, into the physically motivated quantity $\eta$ (equation \ref{eq:equivalent}) to define the identification 
criterion $\Delta \eta>100$ (discussed in Sects. \ref{sec:branch} and \ref{sec:SRandC}). 
By this deeper analysis of the
physical properties of the branches of the binary tree, the Blooming Tree Algorithm can identify
structures with largely different velocity dispersions, as shown by the solid histograms in Figure \ref{fig:vhist}. 
This feature does not belong to the  $\sigma$ plateau algorithm, as shown in Figure 14 of \citet{2015Yu}. 

\section{Conclusion}
\label{sec:discussion}

We present the Blooming Tree Algorithm, a new algorithm for the detection of cluster
substructures and surrounding groups with optical spectroscopic data. The Blooming Tree Algorithm 
is a substantial improvement over our previous procedure, 
the $\sigma$ plateau algorithm, whose performance is described in \citet{2015Yu}.
 
Both algorithms first arrange the galaxies in the field of view in a binary tree according 
to the estimate of the binding energy of each galaxy pair, and search for structures 
associated to the individual branches of the tree. Neither algorithm thus requires an assumption
on the geometry of the systems, on their velocity field or on their dynamical state or an initial
guess of the position and size of the structures to identify.

 The Blooming Tree Algorithm improves over the $\sigma$ plateau algorithm, because it searches all the branches of
the binary tree and relies on a physically motivated combination of velocity dispersion, richness, and size of the candidate structures
to identify them. 
Unlike the $\sigma$ plateau algorithm, that only searches the main branch alone and relies on the velocity dispersion alone, these improvements enable the Blooming Tree Algorithm to identify structures with widely different velocity dispersions and to increase substantially the efficiency of the structure identification. It also enables the Blooming Tree Algorithm to identify easily the galaxy groups present in the cluster outskirts. This ability is relevant for the quantitative investigation of the merging and accretion history of galaxy
clusters \citep{2001Rines,2013Lemze,2016DeBoni}.

Because both methods are based on the arrangement of the galaxies in a hierarchical 
binary tree, the identified structures are naturally nested into each other and 
by walking on the tree branches we can actually separate each individual structure into
smaller and smaller dynamically distinct substrucures.
The binary tree also easily returns a list of the members of the identified structures,
a piece of information that is necessary to estimate the properties of the structures, like
velocity dispersion, size, and mass. 

We test the Blooming Tree Algorithm on 300 mock redshift surveys of 100 clusters of mass $\sim 5\times 10^{14} h^{-1} M_\odot$ 
extracted from an $N$-body simulation of a $\Lambda$CDM model. We consider 50 {\it merging} clusters, whose
most massive substructure other than the cluster core has mass at least half the mass of the core, which is the 
substructure whose center coincides with the cluster center; we also consider 50 {\it normal} clusters where
that condition is not verified.
We also explore mock surveys of different density, by varying the number $N_c$ of galaxies within a sphere of radius 
$6 h^{-1}$~Mpc from the cluster center in the range $N_c=[50-300]$. 
We only consider substructures and surrounding groups with mass larger than $ 10^{13} h^{-1} M_\odot$. 

A substantial fraction of the 3D structures are correctly identified by the Blooming Tree Algorithm. 
Disregarding the case $N_c=50$, that returns too sparse redshift surveys, the completeness of 
the substructure catalogues is $\sim 80\%$ 
for both the normal and merging clusters.
A large completeness is also obtained for the surrounding groups, with $\sim 50\%$ and
$\sim 60\%$ for the normal and merging clusters, respectively. 

The completeness is almost
independent of $N_c$, provided that $N_c>50$. The density of the redshift survey affects the number 
of spurious structures, because with a larger number of galaxies in the field of view, the probability 
of detecting false structures increases. In fact, when increasing $N_c$ from $50$ to $300$, the success rate, 
the fraction of 2D structures that correspond to 
real 3D structures, drops from $75\%$ to $50\%$ for the normal clusters, and from $63\%$ to $49\%$ for the merging clusters. 

These results are rather impressive, because the clusters are extracted from the $N$-body simulation without any particular
criterion in addition to the mass of the cluster and of the largest substructure.

When considering substructures, surrounding groups and cluster cores, the Blooming Tree Algorithm has an overall completeness of $\sim 60\%$ and a success rate in the range $\sim 50-70\%$, substantially 
larger than the completeness and success rate of the $\sigma$ plateau algorithm. 
This latter algorithm has a completeness $\sim 30-50$\% and a success rate $\sim 15-20$\%, depending on the mass and the dynamical state of the cluster \citep{2015Yu}. 

Clearly, for each galaxy, we can only measure three phase-space coordinates out of six
and the estimate of the binding energy of each pairs of galaxies, on which the Blooming Tree Algorithm  is based, 
can heavily be affected by projection effects. In addition, the galaxy peculiar velocities are unknown, 
and the redshift difference entirely contributes to the estimated kinetic energy of the galaxy pair. 
Nevertheless,
the good performance of the Blooming Tree Algorithm shows that these limitations 
little affect the identification procedure and that this algorithm is a powerful tool
to identify the substructures of clusters and their surrounding groups  when 
dense redshift surveys of clusters, like CIRS \citep{Rines2006} and HeCS \citep{Rines2013},
are available. The Blooming Tree Algorithm 
can thus be a powerful tool to infer the merging history of clusters, investigate
their dynamics and constrain their formation models \citep{2016Yu}.

\acknowledgments
AD thanks Margaret Geller for fruitful discussions and suggestions. 
We sincerely thank an anonymous referee whose comments and suggestions urged us to
improve substantially the presentation of our results.
We acknowledge support from the grant I$@$UNITO of the Italian Ministry of Education, 
University and Research assigned to the University of Torino, 
the INFN grant InDark, the National Natural Science Foundation of China under Grants Nos. 11403002, 
the Fundamental Research Funds for the Central Universities and 
Scientific Research Foundation of Beijing Normal University. 
MB also acknowledges the financial contribution by the PRIN INAF 2012 
``The Universe in the box: multiscale simulations of cosmic structure'',
and the support from the Italian Ministry for Education, University and Research (MIUR) 
through the SIR individual grant SIMCODE, project number RBSI14P4IH.

\bibliography{mergecl}

\begin{thebibliography}{52}
\expandafter\ifx\csname natexlab\endcsname\relax\def\natexlab#1{#1}\fi

\bibitem[{{Aguerri} \& {S{\'a}nchez-Janssen}(2010)}]{2010Aguerri}
{Aguerri}, J.~A.~L., \& {S{\'a}nchez-Janssen}, R. 2010, \aap, 521, A28

\bibitem[{{Agulli} {et~al.}(2016){Agulli}, {Aguerri}, {S{\'a}nchez-Janssen},
  {Dalla Vecchia}, {Diaferio}, {Barrena}, {Dominguez Palmero}, \&
  {Yu}}]{2016Agulli}
{Agulli}, I., {Aguerri}, J.~A.~L., {S{\'a}nchez-Janssen}, R., {Dalla Vecchia},
  C., {Diaferio}, A., {Barrena}, R., {Dominguez Palmero}, L., \& {Yu}, H. 2016,
  \mnras, 458, 1590

\bibitem[{{Audit} {et~al.}(1998){Audit}, {Teyssier}, \& {Alimi}}]{1998Audit}
{Audit}, E., {Teyssier}, R., \& {Alimi}, J.-M. 1998, \aap, 333, 779

\bibitem[{{Baldi}(2012)}]{2012Baldi}
{Baldi}, M. 2012, \mnras, 422, 1028

\bibitem[{{Barmby} \& {Huchra}(1998)}]{1998Barmby}
{Barmby}, P., \& {Huchra}, J.~P. 1998, \aj, 115, 6

\bibitem[{{Bird}(1994)}]{1994Bird}
{Bird}, C. 1994, \apj, 422, 480

\bibitem[{{Colberg} {et~al.}(2005){Colberg}, {Krughoff}, \&
  {Connolly}}]{2005Colberg}
{Colberg}, J.~M., {Krughoff}, K.~S., \& {Connolly}, A.~J. 2005, \mnras, 359,
  272

\bibitem[{{Colberg} {et~al.}(1999){Colberg}, {White}, {Jenkins}, \&
  {Pearce}}]{1999Colberg}
{Colberg}, J.~M., {White}, S.~D.~M., {Jenkins}, A., \& {Pearce}, F.~R. 1999,
  \mnras, 308, 593

\bibitem[{{Colless} \& {Dunn}(1996)}]{1996Colless}
{Colless}, M., \& {Dunn}, A.~M. 1996, \apj, 458, 435

\bibitem[{{Davis} {et~al.}(1985){Davis}, {Efstathiou}, {Frenk}, \&
  {White}}]{1985Davis}
{Davis}, M., {Efstathiou}, G., {Frenk}, C.~S., \& {White}, S.~D.~M. 1985, \apj,
  292, 371

\bibitem[{{De Boni} {et~al.}(2016){De Boni}, {Serra}, {Diaferio}, {Giocoli}, \&
  {Baldi}}]{2016DeBoni}
{De Boni}, C., {Serra}, A.~L., {Diaferio}, A., {Giocoli}, C., \& {Baldi}, M.
  2016, \apj, 818, 188

\bibitem[{{Diaferio}(1999)}]{1999Diaferio}
{Diaferio}, A. 1999, \mnras, 309, 610

\bibitem[{{Diaferio} \& {Geller}(1997)}]{1997Diaferio}
{Diaferio}, A., \& {Geller}, M.~J. 1997, \apj, 481, 633

\bibitem[{{Diaferio} {et~al.}(2001){Diaferio}, {Kauffmann}, {Balogh}, {White},
  {Schade}, \& {Ellingson}}]{2001Diaferio}
{Diaferio}, A., {Kauffmann}, G., {Balogh}, M.~L., {White}, S.~D.~M., {Schade},
  D., \& {Ellingson}, E. 2001, \mnras, 323, 999

\bibitem[{{Diemand} {et~al.}(2004){Diemand}, {Moore}, \&
  {Stadel}}]{2004Diemand}
{Diemand}, J., {Moore}, B., \& {Stadel}, J. 2004, \mnras, 352, 535

\bibitem[{{Dressler} {et~al.}(2013){Dressler}, {Oemler}, {Poggianti},
  {Gladders}, {Abramson}, \& {Vulcani}}]{Dressler2013}
{Dressler}, A., {Oemler}, Jr., A., {Poggianti}, B.~M., {Gladders}, M.~D.,
  {Abramson}, L., \& {Vulcani}, B. 2013, \apj, 770, 62

\bibitem[{{Dressler} \& {Shectman}(1988)}]{1988DS}
{Dressler}, A., \& {Shectman}, S.~A. 1988, \aj, 95, 985

\bibitem[{Everitt {et~al.}(2011)Everitt, Landau, Leese, \& Stahl}]{everitt2011}
Everitt, B.~S., Landau, S., Leese, M., \& Stahl, D. 2011, Cluster Analysis, 5th
  Edition, 71

\bibitem[{{Geller} \& {Beers}(1982)}]{1982Geller}
{Geller}, M.~J., \& {Beers}, T.~C. 1982, \pasp, 94, 421

\bibitem[{{Gill} {et~al.}(2005){Gill}, {Knebe}, \& {Gibson}}]{2005Gill}
{Gill}, S.~P.~D., {Knebe}, A., \& {Gibson}, B.~K. 2005, \mnras, 356, 1327

\bibitem[{{Gill} {et~al.}(2004){Gill}, {Knebe}, {Gibson}, \&
  {Dopita}}]{2004Gill}
{Gill}, S.~P.~D., {Knebe}, A., {Gibson}, B.~K., \& {Dopita}, M.~A. 2004,
  \mnras, 351, 410

\bibitem[{{Harvey} {et~al.}(2015){Harvey}, {Massey}, {Kitching}, {Taylor}, \&
  {Tittley}}]{2015Harvey}
{Harvey}, D., {Massey}, R., {Kitching}, T., {Taylor}, A., \& {Tittley}, E.
  2015, Science, 347, 1462

\bibitem[{Hennig {et~al.}(2015)Hennig, Meila, Murtagh, \& Rocci}]{hennig2015}
Hennig, C., Meila, M., Murtagh, F., \& Rocci, R. 2015, Handbook of cluster
  analysis (CRC Press)

\bibitem[{{Huchra} \& {Geller}(1982)}]{1982FoF}
{Huchra}, J.~P., \& {Geller}, M.~J. 1982, \apj, 257, 423

\bibitem[{{Hwang} {et~al.}(2012){Hwang}, {Geller}, {Diaferio}, \&
  {Rines}}]{2012Hwang}
{Hwang}, H.~S., {Geller}, M.~J., {Diaferio}, A., \& {Rines}, K.~J. 2012, \apj,
  752, 64

\bibitem[{{Knebe} \& {M{\"u}ller}(2000)}]{2000Knebe}
{Knebe}, A., \& {M{\"u}ller}, V. 2000, \aap, 354, 761

\bibitem[{{Knebe} {et~al.}(2013){Knebe}, {Pearce}, {Lux}, {Ascasibar},
  {Behroozi}, {Casado}, {Moran}, {Diemand}, {Dolag}, {Dominguez-Tenreiro},
  {Elahi}, {Falck}, {Gottl{\"o}ber}, {Han}, {Klypin}, {Luki{\'c}},
  {Maciejewski}, {McBride}, {Merch{\'a}n}, {Muldrew}, {Neyrinck}, {Onions},
  {Planelles}, {Potter}, {Quilis}, {Rasera}, {Ricker}, {Roy}, {Ruiz},
  {Sgr{\'o}}, {Springel}, {Stadel}, {Sutter}, {Tweed}, \& {Zemp}}]{2013Knebe}
{Knebe}, A. {et~al.} 2013, \mnras, 435, 1618

\bibitem[{{Komatsu} {et~al.}(2011){Komatsu}, {Smith}, {Dunkley}, {Bennett},
  {Gold}, {Hinshaw}, {Jarosik}, {Larson}, {Nolta}, {Page}, {Spergel},
  {Halpern}, {Hill}, {Kogut}, {Limon}, {Meyer}, {Odegard}, {Tucker}, {Weiland},
  {Wollack}, \& {Wright}}]{2011WMAP}
{Komatsu}, E. {et~al.} 2011, \apjs, 192, 18

\bibitem[{{Kummer} {et~al.}(2018){Kummer}, {Kahlhoefer}, \&
  {Schmidt-Hoberg}}]{2017Kummer}
{Kummer}, J., {Kahlhoefer}, F., \& {Schmidt-Hoberg}, K. 2018, \mnras, 474, 388

\bibitem[{{Lemze} {et~al.}(2013){Lemze}, {Postman}, {Genel}, {Ford},
  {Balestra}, {Donahue}, {Kelson}, {Nonino}, {Mercurio}, {Biviano}, {Rosati},
  {Umetsu}, {Sand}, {Koekemoer}, {Meneghetti}, {Melchior}, {Newman}, {Bhatti},
  {Voit}, {Medezinski}, {Zitrin}, {Zheng}, {Broadhurst}, {Bartelmann},
  {Benitez}, {Bouwens}, {Bradley}, {Coe}, {Graves}, {Grillo}, {Infante},
  {Jimenez-Teja}, {Jouvel}, {Lahav}, {Maoz}, {Merten}, {Molino}, {Moustakas},
  {Moustakas}, {Ogaz}, {Scodeggio}, \& {Seitz}}]{2013Lemze}
{Lemze}, D. {et~al.} 2013, \apj, 776, 91

\bibitem[{{Materne}(1978)}]{1978Materne}
{Materne}, J. 1978, \aap, 63, 401

\bibitem[{{Mohammed} {et~al.}(2016){Mohammed}, {Saha}, {Williams},
  {Liesenborgs}, \& {Sebesta}}]{2016Mohammed}
{Mohammed}, I., {Saha}, P., {Williams}, L.~L.~R., {Liesenborgs}, J., \&
  {Sebesta}, K. 2016, \mnras, 459, 1698

\bibitem[{{Mohr} {et~al.}(1996){Mohr}, {Geller}, \& {Wegner}}]{1996Mohr}
{Mohr}, J.~J., {Geller}, M.~J., \& {Wegner}, G. 1996, \aj, 112, 1816

\bibitem[{{Natarajan} {et~al.}(2007){Natarajan}, {De Lucia}, \&
  {Springel}}]{2007Natarajan}
{Natarajan}, P., {De Lucia}, G., \& {Springel}, V. 2007, \mnras, 376, 180

\bibitem[{{Oguri} {et~al.}(2018){Oguri}, {Miyazaki}, {Hikage}, {Mandelbaum},
  {Utsumi}, {Miyatake}, {Takada}, {Armstrong}, {Bosch}, {Komiyama},
  {Leauthaud}, {More}, {Nishizawa}, {Okabe}, \& {Tanaka}}]{2017Oguri}
{Oguri}, M. {et~al.} 2018, \pasj, 70, S26

\bibitem[{{Okabe} {et~al.}(2014){Okabe}, {Futamase}, {Kajisawa}, \&
  {Kuroshima}}]{2014Okabe}
{Okabe}, N., {Futamase}, T., {Kajisawa}, M., \& {Kuroshima}, R. 2014, \apj,
  784, 90

\bibitem[{{Pisani}(1996)}]{1996Pisani}
{Pisani}, A. 1996, \mnras, 278, 697

\bibitem[{{Pranger} {et~al.}(2013){Pranger}, {B{\"o}hm}, {Ferrari}, {Diaferio},
  {Hunstead}, {Maurogordato}, {Benoist}, {Brinchmann}, \&
  {Schindler}}]{2013Pranger}
{Pranger}, F. {et~al.} 2013, \aap, 557, A62

\bibitem[{{Ramella} {et~al.}(2007){Ramella}, {Biviano}, {Pisani}, {Varela},
  {Bettoni}, {Couch}, {D'Onofrio}, {Dressler}, {Fasano}, {Kj{\o}rgaard},
  {Moles}, {Pignatelli}, \& {Poggianti}}]{2007Ramella}
{Ramella}, M. {et~al.} 2007, \aap, 470, 39

\bibitem[{{Rines} \& {Diaferio}(2006)}]{Rines2006}
{Rines}, K., \& {Diaferio}, A. 2006, \aj, 132, 1275

\bibitem[{{Rines} {et~al.}(2008){Rines}, {Diaferio}, \&
  {Natarajan}}]{2008Rines}
{Rines}, K., {Diaferio}, A., \& {Natarajan}, P. 2008, \apjl, 679, L1

\bibitem[{{Rines} {et~al.}(2013){Rines}, {Geller}, {Diaferio}, \&
  {Kurtz}}]{Rines2013}
{Rines}, K., {Geller}, M.~J., {Diaferio}, A., \& {Kurtz}, M.~J. 2013, \apj,
  767, 15

\bibitem[{{Rines} {et~al.}(2001){Rines}, {Mahdavi}, {Geller}, {Diaferio},
  {Mohr}, \& {Wegner}}]{2001Rines}
{Rines}, K., {Mahdavi}, A., {Geller}, M.~J., {Diaferio}, A., {Mohr}, J.~J., \&
  {Wegner}, G. 2001, \apj, 555, 558

\bibitem[{{Rines} {et~al.}(2016){Rines}, {Geller}, {Diaferio}, \&
  {Hwang}}]{2016Rines}
{Rines}, K.~J., {Geller}, M.~J., {Diaferio}, A., \& {Hwang}, H.~S. 2016, \apj,
  819, 63

\bibitem[{{Robertson} {et~al.}(2017){Robertson}, {Massey}, \&
  {Eke}}]{2017Robertson}
{Robertson}, A., {Massey}, R., \& {Eke}, V. 2017, \mnras, 465, 569

\bibitem[{{Serna} \& {Gerbal}(1996)}]{1996Serna}
{Serna}, A., \& {Gerbal}, D. 1996, \aap, 309, 65

\bibitem[{{Serra} {et~al.}(2011){Serra}, {Diaferio}, {Murante}, \&
  {Borgani}}]{Serra2011}
{Serra}, A.~L., {Diaferio}, A., {Murante}, G., \& {Borgani}, S. 2011, \mnras,
  412, 800

\bibitem[{{Solanes} {et~al.}(1999){Solanes}, {Salvador-Sol{\'e}}, \&
  {Gonz{\'a}lez-Casado}}]{1999Solanes}
{Solanes}, J.~M., {Salvador-Sol{\'e}}, E., \& {Gonz{\'a}lez-Casado}, G. 1999,
  \aap, 343, 733

\bibitem[{{Springel} {et~al.}(2001){Springel}, {White}, {Tormen}, \&
  {Kauffmann}}]{2001Springel}
{Springel}, V., {White}, S.~D.~M., {Tormen}, G., \& {Kauffmann}, G. 2001,
  \mnras, 328, 726

\bibitem[{{Utsumi} {et~al.}(2016){Utsumi}, {Geller}, {Dell'Antonio}, {Kamata},
  {Kawanomoto}, {Koike}, {Komiyama}, {Koshida}, {Mineo}, {Miyazaki}, {Sakurai},
  {Tait}, {Terai}, {Tomono}, {Usuda}, {Yamada}, \& {Zahid}}]{2016Utsumi}
{Utsumi}, Y. {et~al.} 2016, \apj, 833, 156

\bibitem[{{Yu} {et~al.}(2016){Yu}, {Diaferio}, {Agulli}, {Aguerri}, \&
  {Tozzi}}]{2016Yu}
{Yu}, H., {Diaferio}, A., {Agulli}, I., {Aguerri}, J.~A.~L., \& {Tozzi}, P.
  2016, \apj, 831, 156

\bibitem[{{Yu} {et~al.}(2015){Yu}, {Serra}, {Diaferio}, \& {Baldi}}]{2015Yu}
{Yu}, H., {Serra}, A.~L., {Diaferio}, A., \& {Baldi}, M. 2015, \apj, 810, 37

\end{thebibliography}

\end{CJK*}
\end{document}